\documentclass[aps,prx, twocolumn,amsmath, longbibliography, amssymb,superscriptaddress]{revtex4-1}

\usepackage{bm}
\usepackage[retainorgcmds]{IEEEtrantools}
\usepackage{graphicx}
\usepackage{mathrsfs}
\usepackage{amsmath}
\usepackage{amssymb}
\usepackage{color}
\usepackage{amsfonts}
\usepackage{times,txfonts}
\usepackage{nicefrac}
\usepackage[colorlinks=true,linkcolor=blue,urlcolor=blue,citecolor=blue,pdfusetitle]{hyperref}
\usepackage{physics}

\newcommand{\scnd}{2^{\text{nd}}}

\begin{document}

%\title{Thermodynamics of continuous quantum measurements: \\[0.1cm] A collisional model approach}
\title{Informational steady-states and conditional entropy production in continuously monitored systems}
\date{\today}
\author{Gabriel T. Landi}
\email{gtlandi@if.usp.br}
\affiliation{Instituto de F\'isica da Universidade de S\~ao Paulo,  05314-970 S\~ao Paulo, Brazil.}
\author{Mauro Paternostro}
\affiliation{Centre for Theoretical Atomic, Molecular, and Optical Physics, School of Mathematics and Physics, Queens University, Belfast BT7 1NN, United Kingdom}
\author{Alessio Belenchia}
\affiliation{Institut f\"{u}r Theoretische Physik, Eberhard-Karls-Universit\"{a}t T\"{u}bingen, 72076 T\"{u}bingen, Germany}
\affiliation{Centre for Theoretical Atomic, Molecular, and Optical Physics, School of Mathematics and Physics, Queens University, Belfast BT7 1NN, United Kingdom}

\begin{abstract}
%The degree of irreversibility of an open system dynamics can be quantified by the entropy production, the key quantity behind the $\scnd$ law of thermodynamics. Part of this irreversibility, however, is informational, being associated to our ignorance about the state of the environment. Access to even part of the environment provides information that would reduces the irreversibility of the process as quantified by a conditional entropy production. In the case of systems which are continuously monitored, such process allows for the introduction of \emph{informational steady-states} (ISSs), i.e. stationary states of a conditional dynamics that are maintained owing to the unbroken acquisition of information. 
We put forth a unifying formalism for the description of the thermodynamics of continuously monitored systems, where measurements are only performed on the environment connected to a system. 
%Our framework is  based on collisional models and
%and handles both discrete and continuous variable quantum system. 
We show, in particular, that the conditional and unconditional entropy production, which quantify the degree of irreversibility of the open system's dynamics, are related to each other by the Holevo quantity. %, one of the most fundamental objects in quantum information theory.
This, in turn, can be further split into an information gain rate and loss rate, which provide conditions for the existence of \emph{informational steady-states} (ISSs), i.e. stationary states of a conditional dynamics that are maintained owing to the unbroken acquisition of information. 
We illustrate the applicability of our framework through several examples.%, involving both discrete and continuous-variable systems. We also demonstrate the practical usefulness of our theory by providing a detailed account of a recent optomechanical experiment [Phys. Rev. Lett, {\bf 125}, 080601 (2020)].
\end{abstract}

\maketitle 

%\tableofcontents
%\newpage
%\end{widetext}
%

%%%%%%%%%%%%%%%%%%%%%%%%%%%%%%%%%%%%%%%%%%
%
%
\section{Introduction}
%
%
%%%%%%%%%%%%%%%%%%%%%%%%%%%%%%%%%%%%%%%%%%

The dynamics of a quantum system depends not only on itself, but also on how it is probed, showcasing the remarkable extrinsic character of quantum mechanics. 
This unavoidable backaction due to measurements can be directly probed in the laboratory~\cite{Murch2008,Purdy2013,Teufel2016,Minev2018}, and is by far the most intriguing and dramatic aspect of quantum theory. 
It also has a clear thermodynamic flavor~\cite{Binder2018a}, since backaction is an intrinsically irreversible process. 
A comprehensive theory describing the thermodynamics of monitored systems would therefore greatly benefit our understanding of the interplay between information and dissipation. 
Constructing such a theory, however, is not trivial, since it requires reformulating  the 2nd law to take into account the information learned from the measurements. 
We call this a \emph{conditional $2^\text{nd}$ law}.
It quantifies which processes are allowed, \emph{given} a certain set of measurement outcomes. 
Interestingly, due to measurement backaction, the noise introduced by the measurement can actually make the conditional process more irreversible, as recently demonstrated in a superconducting qubit experiment~\cite{Naghiloo2018}. 
%How to construct a strategy which avoids this altogether is still an open question. 

When a system is coupled to two baths at different temperatures, it  usually tends to a non-equilibrium steady-state (NESS), where the competition between the two baths keeps the system away from equilibrium.
Continuous measurements can lead to a similar effect.
In this case, noise is constantly being introduced by the environment or the measurement backaction. 
But information is also constantly being acquired. 
These two effects compete, leading the system toward an \emph{informational steady-state (ISS)}. 
Crucially, the ISS relies on the experimenter's knowledge of the measurement records. 
%If the measurements are performed, but the outcomes are not read, the steady-state would be solely due to the dissipation of the apparatus. We call this the unconditional NESS (uNESS). 
%But when the outcomes are known,  the inferences we can make about the state of the system are different. This is the conditional NESS (cNESS = ISS). 
A beautiful experimental illustration of this effect was  recently given in~\cite{Rossi2019}, where the authors studied an optomechanical membrane monitored by an optical field. 
By measuring the  field, one could  monitor the position of the mechanical membrane and thus infer a steady-state which was close to the ground state. 
Conversely, if the measurements are not read, the membrane is perceived to be in a thermal state with higher temperatures.
The ISS is therefore \emph{colder}, due to the information acquired from the continuous measurement.

ISSs are just one example of the many interesting phenomena that emerge when quantum measurements are introduced in a thermodynamic picture. 
The deep connections between the two concepts, together with recent experimental advances in controlled quantum platforms, have led to a surge of interest in formulating conditional laws of thermodynamics~\cite{Sagawa2008,Ito2013,Sagawa2012,Sagawa2013,Funo2013,Elouard2017a,Buffoni2018,Mohammady2019a,Beyer2020,Sone2020,Strasberg2019a,Strasberg2020,Belenchia2019}. 
This also motivated ground-breaking experiments applying these ideas to Maxwell demon engines and feedback control~\cite{Toyabe2010,Koski2014,Cottet2017,Naghiloo2018,Debiossac2019}.  
%Often, these constructions involve the identification of quantities that can quantify the gain of information and the measurement backaction, and how these affect the  degree of irreversibility of the process. 
In all these frameworks, however, the measurements are assumed to act directly on the system, making them explicitly invasive.
 
Conversely, our interest in this paper will be on formulating the laws of thermodynamics when the measurements are done only on the environment and \emph{only} after it interacted with the system. 
The scenario is therefore non-invasive by construction, so that any information acquired can only make the process more reversible, even if the measurement is very poor (as is often the case when dealing with large environments).
This represents a change in philosophy compared to, e.g., Ref.~\cite{Funo2013}, where the measurement was introduced by coupling the system to a memory and then measuring the memory.
In that case one constructs the conditional $\scnd$ law by comparing the situation where the system is fully isolated, with that in which it is open due to the interaction with the memory. 
In our case, we assume instead that the interaction between system and bath is inevitable and will happen whether or not we measure it.
We then ask how measuring the bath affects the degree of irreversibility of the process. 
 
Crucially, the framework we develop will focus on continuously monitored system, in contrast to e.g. Ref~\cite{Funo2013}. 
It is therefore particularly suited for describing ISSs.
Our endeavor began in Ref.~\cite{Belenchia2019}, where we put forth a semiclassical theory valid for Gaussian processes. 
We were interested in quantum optical experiments, which have already been using some of these ideas for many decades, in the framework of continuously monitored systems~\cite{Wiseman2009,Jacobs2014}.
In fact, our theory was recently employed in~\cite{Rossi2020} to experimentally assess the conditional $2^\text{nd}$ law in an optomechanical system. 
However, in addition to being semiclassical,  the framework of Ref.~\cite{Belenchia2019} also has another serious limitation: it is formulated solely in terms of the  stochastic master equation obeyed by the system; that is, it does not require an explicit model of the environment, but only which type of open dynamics it produces. 

There has been increasing evidence that a proper formulation of thermodynamics in the quantum regime is only possible if information on the environment and the system-environment interactions are provided~\cite{Landi2020a}. 
Reduced descriptions, based only on master equations, can  show apparent violations of the $\scnd$ law~\cite{Levy2014}, something which can only be resolved by introducing a specific model of the environment~\cite{DeChiara2018}. 
%Our efforts in modeling the experiments in~\cite{Rossi2020} also strongly corroborate this view.
%In fact, an explicit illustration will be provided below, where we will show that models that yield the same type of master equation can have completely different thermodynamics features.

\begin{figure*}
\centering
\includegraphics[width=0.9\textwidth]{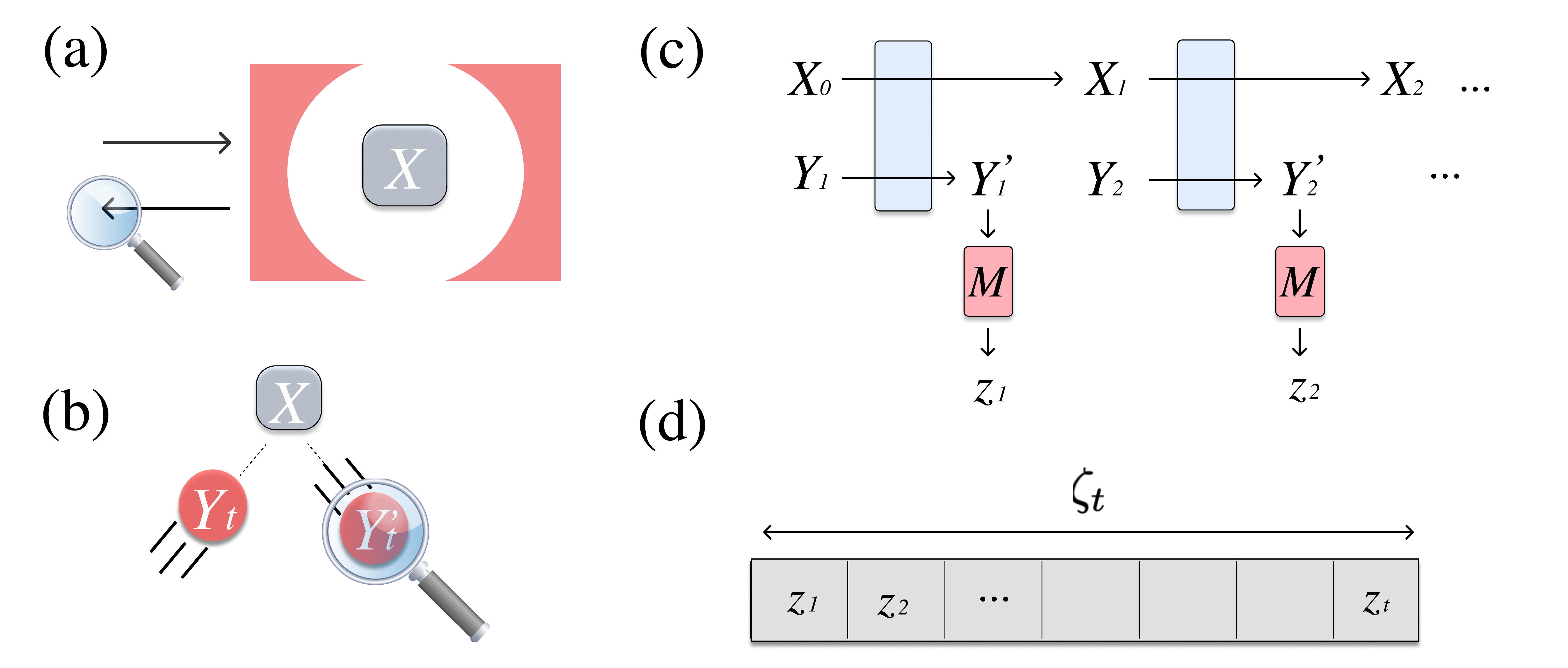}
\caption{\label{fig:drawings}
%Continuously measured collisional models ($\text{CM}^2$). 
(a) A typical method for continuously monitoring a system is to couple it to an optical cavity and measure the photons leaking out.
(b) In a collisional model picture, the monitoring is introduced instead through a series of sequential collisions between the system $X$ and independent ancillae $Y_t$, which are subjected to measurement after each collision. 
(c) Diagrammatic representation of the model. The system is described  stroboscopically (discrete time)  by a state $\rho_{X_t}$. At each instant of time, it interacts with an independent ancilla, prepared in state $\rho_Y$, according to the map in Eq.~\eqref{global_map}. 
Afterwards, the ancillae are measured, as described by generalized measurement operators $\{M_z\}$, which produce a classical (and random) outcome $z_t$. 
(d) As time progresses, one builds up a measurement record $\zeta_t = (z_1,\ldots,z_{t})$, which contains all the information acquired about $X$ up to time $t$. 
}
\end{figure*}

In this paper we put forth a very general framework for describing the thermodynamics of continuously monitored systems, where measurements are only done indirectly in the bath. 
The formalism applies to a broad variety of systems and process, and is particularly suited for describing ISSs. 
The building block we use is to replace the continuous dynamics by a stroboscopic evolution in small time-steps, described in terms of a collisional model (CM)~\cite{Rau1963,Scarani2002,Ziman2002,Englert2002,Attal2006a,Karevski2009,Pellegrini2009,Giovannetti2012,Rybar2012a,Strasberg2016}.
This has two main advantages. 
First, the thermodynamics of CMs is by now very well understood~\cite{Strasberg2016,Rodrigues2019,DeChiara2018,Barra2015,Pereira2018}  (see also~\cite{Landi2020a} for a recent review). 
And second, CMs naturally emerge in quantum optics, from a discretization of the field operator into discrete time-bins~\cite{Ciccarello2017,Gross2017a}. 
The typical scenario is a system interacting with an optical cavity, where a constant flow of photons is injected by an external pump [cf. Fig.~\ref{fig:drawings}(a)]. At each time step, the system will only interact with a certain time-window of the input/output field, thus transforming the dynamics into that of a series of sequential collisions between the system and some ancilla. 
Due to this connection, collisional models serve as a convenient tool for constructing the framework of continuous measurements in experimentally relevant systems. 
We refer to these as Continuously Monitored Collisional Models ($\text{CM}^2$).

Our paper is organized as follows. 
Sec.~\ref{sec:setup} establishes the basic framework, including the collisional setup. The corresponding information flows and thermodynamic features are characterized in Sec.~\ref{sec:info}, which also contains the main contribution of this work: namely the construction of a conditional $\scnd$ law, which is capable of capturing the interplay between thermodynamics and information. In Sec.~\ref{sec:qubits}, we apply the $\text{CM}^2$ framework to models involving qubits providing some illustrative applications.
Accompanying this manuscript, we also make publicly available a self-contained numerical library in Mathematica, for carrying out stochastic simulations of $\text{CM}^2$s \footnote{The code can be downloaded \href{https://www.wolframcloud.com/obj/gtlandi/Published/CM2_qubit_models.nb}{here}.}.
Finally, in Sec.~\ref{sec:conc} we draw our conclusions and highlight the perspectives opened by our approach.

%%%%%%%%%%%%%%%%%%%%%%%%%%%%%%
%
%
\section{\label{sec:setup}Continuously measured collisional models ($\text{CM}^2$)}
%
%
%%%%%%%%%%%%%%%%%%%%%%%%%%%%%%

Here we develop the basic framework of $\text{CM}^2$. 
We consider a system $X$, with initial density matrix $\rho_{X_0}$, which is put to interact sequentially with a series of independent and identically prepared (iid) ancillae, labelled $Y_1$, $Y_2$ etc., and prepared always in the same state $\rho_{Y_t} = \rho_Y$. 
Time is labeled in discrete units of $t = 0,1,2,3,\ldots$. 
The collision taking the system from ${t-1}$ to ${t}$ is described by a unitary $U_{t}$ acting only between the system $X$ and ancilla $Y_{t}$ as (Fig.~\ref{fig:drawings}(b)):
\begin{equation}\label{global_map}
\rho_{X_{t}Y_{t}'} = U_t(\rho_{X_{t-1}} \otimes \rho_{Y_t}) U_t^\dagger, 
\end{equation}
where $Y_t'$ refers to the state of ancilla $Y_t$ after the collision.
Taking the partial trace over the ancilla leads to the stroboscopic (Markovian) map  
\begin{equation}\label{unconditional_map}
\rho_{X_{t}}  = \mathcal{E}(\rho_{X_{t-1}}) := \tr_{Y_t} \left\{\rho_{X_{t}Y_{t}'}\right\}.%\Big\{ U_t(\rho_{X_{t-1}} \otimes \rho_{Y_t}) U_t^\dagger \Big\}.
\end{equation}
Notice that $\mathcal{E}$ does not need to carry an index $t$, since it is the same for all collisions. 
After such map, the ancilla $Y_t'$ never participates again in the dynamics and, for the next step, a fresh ancilla $Y_{t+1}$ is introduced and the map in  Eq.~\eqref{unconditional_map} is repeated.

Information on the state of the system is acquired indirectly by measuring the states $\rho_{Y_t}'$ of each ancilla after they collided with $X$.
The measurement is described by a set of generalized measurement operators $\{M_z\}$, satisfying $\sum_z M_z^\dagger M_z = \openone$, so that outcome $z_t$ occurs with probability 
\begin{equation}\label{Pz}
P(z_t) = \tr\Big\{ M_{z_t} \rho_{Y_t}'  M_{z_t}^\dagger \Big\}.
\end{equation}
By using generalized measurements, we  encompass both projective, as well as weak measurements in the bath.
A diagrammatic depiction of the dynamics is shown in Fig.~\ref{fig:drawings}(c).
A $\text{CM}^2$ is completely described by specifying $\{ \rho_Y, U, M_z\}$.

The distribution in Eq.~\eqref{Pz} concerns only the marginal statistics of a single outcome. 
Our interest will be instead on the \emph{joint statistics} of the set of measurement records 
\begin{equation}\label{zeta}
\zeta_t = (z_1, \ldots, z_t). 
\end{equation}
The indices are chosen so that $\zeta_t$ contains all information about the system available up to time $t$. 
As $\zeta$ encompasses the entire measurement record, it is associated with the ``integrated'' information on $X$. Conversely, $z_t$ represents a \emph{differential information gain} associated only with the step $X_{t-1} \to X_{t}$ (Fig.~\ref{fig:drawings}(d)). 
The joint distribution $P(\zeta_t)$ is given by 
\begin{equation}
P(\zeta_t) = \tr_{XY_1\ldots Y_{t}} \Big\{ M_{z_{t}} \ldots M_{z_1} \rho_{XY_1\ldots Y_{t}} M_{z_1}^\dagger \ldots M_{z_{t}}^\dagger \Big\}, 
\end{equation}
where 
\begin{equation*}
\rho_{XY_1\ldots Y_{t}} = \left(\Pi^t_{k=1}U_{k}\right)\left( \rho_{X_0} \bigotimes^t_{j=1} \rho_{Y_j} \right)\left( \Pi^t_{k=1}U_{k}\right)^\dagger. 
\end{equation*}
Note that since the measurements act only on those ancillae that no longer participate in the dynamics, it is irrelevant whether the measurement $M_{z_t}$ occurs before the next evolution with $Y_{t+1}$ or not. 

Finally, we also require the \emph{conditional state} of the system $\rho_{X_t|\zeta_t}$, which quantifies the knowledge the experimenter has about the system, {given} that the measurement record $\zeta_t$ was observed. 
Such state is given by 
\begin{equation}\label{conditional_state}
\rho_{X_t|\zeta_t} = \frac{1}{P(\zeta_t)} \tr_{Y_1\ldots Y_{t}} \Big\{ 
\left(\Pi^t_{k=1}M_{z_k}\right) \rho_{XY_1\ldots Y_{t}} 
\left(\Pi^t_{k=1}M_{z_k}\right)^\dagger \Big\}.
\end{equation}
As the measurements are performed only on the ancillae, there is never a direct backaction on the system, which is expressed mathematically by
\begin{equation}\label{marginal_rho_Xs}
\sum\limits_{\zeta_t} P(\zeta_t) \rho_{X_t | \zeta_t} = \rho_{X_t}
\end{equation}
for any choice of generalized measurements $\{M_z\}$. 
That is, the average of $\rho_{X_t|\zeta_t}$ over all outcomes $\zeta_t$ yields back the unconditional state $\rho_{X_t}$. 
Thus, while there may be a conditional backaction, unconditionally the measurement is non-invasive.

The normalization factor $P(\zeta_t)$ in Eq.~\eqref{conditional_state} introduces a unwanted complication, as it
forbids us to write $\rho_{X_t|\zeta_t}$ as a map acting on $\rho_{X_{t-1} | \zeta_{t-1}}$. 
This can be resolved, however, if we work with unnormalized states. 
We define the completely positive, trace non-preserving map
\begin{equation}\label{conditional_map_def}
\mathcal{E}_z(\rho_X) = \tr_Y \Big\{ M_z U (\rho_X \otimes \rho_Y) U^\dagger M_z^\dagger \Big\}, 
\end{equation}
which is indexed by the possible outcomes $z$ of the measurements. 
Instead of working with $\rho_{X_t|\zeta_t}$ in Eq.~\eqref{conditional_state}, we consider  the unnormalized states $\varrho_{X_t|\zeta_t}$, defined as the sequence generated by the map 
\begin{equation}\label{conditional_map}
\varrho_{X_{t} | \zeta_{t} } = \mathcal{E}_{z_t} \big( \varrho_{X_{t-1}|\zeta_{t-1}}\big)
\end{equation}
with initial condition $\varrho_{X_0|\zeta_0} = \rho_{X_0}$. One may readily verify that 
\begin{equation}\label{normalization_Pzeta}
\tr_X \varrho_{X_t|\zeta_t} = \tr_X \Big\{ \mathcal{E}_{z_{t}} \circ \ldots \circ \mathcal{E}_{z_1} (\rho_{X_0}) \Big\} = P(\zeta_t). 
\end{equation}
The states $\varrho_{X_t|\zeta_t}$ therefore contain the outcome distribution $P(\zeta_t)$ at any given time.
And the normalized state in~\eqref{conditional_state} is recovered as $\rho_{X_t|\zeta_t} = \varrho_{X_t|\zeta_t} / P(\zeta_t)$.

It is useful to keep in mind the  interpretation of a $\text{CM}^2$ as a Hidden Markov model~\cite{Darwiche2009,Neapolitan2003,Ito2013}. 
The system evolution is Markovian, but  this is hidden from the observer who is partially ignorant about its dynamics: access to $X$  is only possible through the classical outcomes $\zeta_t$.
In the language of Bayesian networks, the key issue entailed by our framework is thus about the predictions that can be made on the state of the hidden layer  $X$ given the information available through the visible layer of the outcomes $\zeta_t$ only. 
This  highlights the nice interplay between quantum and classical features, present in these models:
The evolution of the system is quantum but information is only accessed through classical data.
We have also found it illuminating to understand what would be the classical version of a $\text{CM}^2$, as this allows us to relate our framework directly with the classical formalism of Ito, Sagawa and Ueda~\cite{Sagawa2012,Ito2013}.
This  is addressed in Appendix~\ref{app:incoherent_models}, where we also discuss the conditions for a $\text{CM}^2$ to be incoherent.

%%%%%%%%%%%%%%%%%%%%%%%%%%%%%%
%
%
\section{\label{sec:info} Information and thermodynamics}
%
%
%%%%%%%%%%%%%%%%%%%%%%%%%%%%%%

%%%%%%%%%%%%%%%%%%%%%%%%%%%%%%
\subsection{Quantum-classical information}

The information content in the unconditional state $\rho_{X_t}$ can be quantified  by the von Neumann entropy $S(X_t) \equiv S(\rho_{X_t}) = - \tr \rho_{X_t} \ln \rho_{X_t}$. 
Similarly, the information in the conditional state $\rho_{X_t|\zeta_t}$ (properly normalized) is quantified by 
\emph{quantum-classical} conditional entropy
\begin{equation}\label{condS}
S(X_t | \zeta_t) = \sum\limits_{\zeta_t} P(\zeta_t) S(\rho_{X_t|\zeta_t}). 
\end{equation}
Each term $S(\rho_{X_t|\zeta_t})$ quantifies the information for one specific realization $\zeta_t$, and 
$S(X_t | \zeta_t)$ is then an average over all trajectories. 
Note also that this is not the quantum conditional entropy, a quantity which can be negative. 
Here, since we are conditioning on classical outcomes, $S(X_t | \zeta_t)$ is always strictly non-negative.
In this paper all conditional entropies will be of this form. 

The mismatch between $S(X_t)$ and $S(X_t|\zeta_t)$ is given by the \emph{Holevo information} (or Holevo quantity)~\cite{Holevo1973} 
\begin{equation}\label{holevo}
I(X_t \! :\! \zeta_t) := S(X_t) - S(X_t | \zeta_t). 
\end{equation}
It quantifies the information about $X$ contained in the classical outcomes $\zeta_t$.
Its interpretation becomes clearer by casting it as 
\begin{equation}\label{holevo_KL}
I(X_t \! :\! \zeta_t)= \sum\limits_{\zeta_t} P(\zeta_t) \;D\big( \rho_{X_t|\zeta_t} || \rho_{X_t}\big) \geqslant 0, 
\end{equation}
where $D(\rho||\sigma) = \tr(\rho \ln \rho - \rho \ln \sigma)$ is the quantum relative entropy. 
Therefore, $I(X_t \! :\! \zeta_t)$ is the weighted average of the ``distance'' between $\rho_{X_t|\zeta_t}$ and $\rho_{X_t}$.

The Holevo information reflects the \emph{integrated} information, acquired about the system, up to time $t$. 
This is different from the small increment that is obtained from a single outcome $z$, at each step. 
In order to quantify such \emph{differential} information gain, the natural quantity is the conditional 
Holevo information 
\begin{IEEEeqnarray}{rCl}\label{gain}
G_t := I_c(X_{t}\!:\! z_{t}| \zeta_{t-1}) 
&=& I(X_{t}\!:\! \zeta_{t}) - I(X_{t}\!:\! \zeta_{t-1}) \\[0.2cm]
&=&  S(X_t|\zeta_{t-1}) - S(X_t | \zeta_t).
\nonumber
\end{IEEEeqnarray}
It describes the correlations between $X_t$ and the latest available outcome $z_{t}$, given the past outcomes $\zeta_{t-1} = (z_1,\ldots,z_{t-1})$.
The first term involves the state $\rho_{X_t|\zeta_{t-1}}$, which stands for the state of the system at time $t$, conditioned on all measurement records, except the last one.
In symbols, it can thus be written as
\begin{equation}\label{intermediate_unconditional_state}
\rho_{X_t|\zeta_{t-1}} = \mathcal{E}(\rho_{X_{t-1}|\zeta_{t-1}}), 
\end{equation}
where $\mathcal{E}$ is the unconditional map in Eq.~\eqref{unconditional_map}.
%That is, one evolves conditionally up to $t-1$; but the last step is unconditional. 
This therefore affords a beautiful interpretation to 
Eq.~\eqref{gain}.
Starting at $\rho_{X_{t-1}|\zeta_{t-1}}$, one compares two paths: 
a conditional evolution taking $\rho_{X_{t-1}|\zeta_{t-1}} \to \rho_{X_t|\zeta_t}$ and a unconditional evolution taking 
$\rho_{X_{t-1}|\zeta_{t-1}} \to \rho_{X_t|\zeta_{t-1}}$.
Eq.~\eqref{gain} measures the gain in information of the latter, compared to the former.

%At this point it is worth contrasting the Holevo quantity with the so-called Groenewold-Ozawa quantum-classical information~\cite{Groenewold1971,Ozawa1986}, which appears in related works~\cite{Funo2013,Naghiloo2018}. 
%In our notation, this quantity would read 
%\begin{equation}\label{GO}
%I_t^{GO} = S(X_{t-1} | \zeta_{t-1}) - S(X_t | \zeta_t). 
%\end{equation}
%Compared with~\eqref{holevo}, it is clear they quantify completely different quantities. 
%Eq.~\eqref{GO} contrasts the amount of information acquired from the measurement 
%

%%%%%%%%%%%%%%%%%%%%%%%%%%%%%%
\subsection{Information rates and informational steady-states}

Eq.~\eqref{holevo} is always non-negative. However, this does not imply that it will necessarily increase with time. 
In fact, the \emph{information rate}
\begin{equation}\label{Delta_I}
\Delta I_t := I(X_t \! :\! \zeta_t) - I(X_{t-1} \! :\! \zeta_{t-1}) %\gtrless 0, 
\end{equation}
can take any sign.
This reflects the trade-off between the gain in information and the measurement backaction. 
%Since the information is extracted by making the system collide with several ancillae, this process also introduces noise. 
%If the noise is too large, the backaction may be larger than the gain, causing $\Delta I_t$ to be negative. 
%In fact, it helps to picture two extreme cases. If the system starts in an extremely noisy state, any information is 
%a plus and so $I(X_t \! :\! \zeta_t)$ will tend to increase. 
%But if the system starts in a highly pure state, the collisions can cause $I(X_t \! :\! \zeta_t)$ to initially grow but eventually go down with time. 
A natural question is then whether it is possible to split $\Delta I_t$ as the difference between two strictly non-negative terms, the first naturally identified with the differential gain of information~\eqref{gain}, and the second to the differential information loss. 
That is, whether a splitting of the form
\begin{equation}\label{Delta_I_splitting}
\Delta I_t= G_t - L_t, 
\end{equation}
would lead to the identification of a loss term $L_t$ which is strictly non-negative. As we will see in what follows, the answer to this question is in the positive.  
%But it turns out it is. 

To find a formula for $L_t$ we simply insert the first line of~\eqref{gain} into Eq.~\eqref{Delta_I} to find
%
%\begin{equation}\label{Delta_I_2}
%\Delta I_t = I_c(X_{t}\!:\! z_{t}| \zeta_{t-1})  + I(X_t \! : \! \zeta_{t-1}) - I(X_{t-1} \! : \! \zeta_{t-1}) 
%\end{equation}
%The first term is the differential information gain~\eqref{gain}, so we naturally recognize 
%\begin{equation}\label{G}
%G_t = I_c(X_{t}\!:\! z_{t}| \zeta_{t-1}).
%\end{equation}
%The last two terms, on the other hand, refer to the change in the system, from $X_{t-1}$ to $X_t$, with fixed $\zeta_{t-1}$.
%And, as we show next, they indeed yield a negative contribution. That is, 
\begin{equation}\label{loss} 
L_t := I(X_{t-1} \! : \! \zeta_{t-1}) -  I(X_t \! : \! \zeta_{t-1}).
\end{equation}
This is already clearly interpretable as a loss term, as it measures how information is degraded by the map in Eq.~\eqref{intermediate_unconditional_state}. 
Indeed, we can show that it is strictly non-negative. 
To do that, we use Eq.~\eqref{holevo_KL} to write $L_t$ as 
\begin{equation}
\label{Lt}
L_t = \sum\limits_{\zeta_{t-1}} P(\zeta_{t-1}) \Big[  
D(\rho_{X_{t-1} | \zeta_{t-1}} || \rho_{X_{t-1}})-D(\rho_{X_t | \zeta_{t-1}} || \rho_{X_t} )  \Big]. 
\end{equation}
But $\rho_{X_t} = \mathcal{E} (\rho_{X_{t-1}})$ [Eq.~\eqref{unconditional_map}] and $\rho_{X_t|\zeta_{t-1}} = \mathcal{E} (\rho_{X_{t-1} | \zeta_{t-1}})$ [Eq.~\eqref{intermediate_unconditional_state}].
Together with % the two relative entropies inside the sum have the form $D(\rho|| \sigma)-D(\mathcal{E}(\rho) || \mathcal{E}(\sigma))$, which, according to 
the data processing inequality~\cite{Nielsen}, this is enough to ascertain the non-negativity of $L_t$ for any quantum channel $\mathcal{E}$. 
%Whence, $L_t \geqslant 0$. 

%Therefore, within our framework the net information about the system is measured by the Holevo information~\eqref{holevo}. As the system evolves, this quantity may either increase or decrease, so that the information rate $\Delta I_t$ in Eq.~\eqref{Delta_I} does not necessarily have a definite sign. Notwithstanding, we can split $\Delta I_t$ as the difference between two non-negative terms, $G_t$ and $L_t$, Eqs.~\eqref{gain} and~\eqref{loss},which can be viewed as the information gain rate and the information loss rate.  

In the long time limit the system may reach a steady-state where $I_\infty$ no longer changes, so $\Delta I_\infty = 0$. 
This does not necessarily mean $G_\infty = L_\infty= 0$, however. 
It might simply stem from a mutual balancing of gains and losses. That is, $G_\infty = L_\infty\neq 0$. 
We  define an informational steady-state (ISS) as the asymptotic state for which 
\begin{equation}\label{ISS}
\Delta I_\text{ISS} = 0 \quad \text{ but} \quad G_\text{ISS} = L_\text{ISS} \neq 0. 
\end{equation}
In an ISS, information is continuously acquired, but this is balanced by the noise that is introduced by the measurement.
Crucially, the ISS does \emph{not} mean that $\rho_{X_t|\zeta_t}$ is no longer changing. 
This state is stochastic and thus continues to evolve indefinitely. Instead, what become stationary is the stochastic distribution of states in state-space~\cite{Ficheux2018}.
%\footnote{An explicit example is shown in Fig.~\ref{fig:time_series}(h)}.

%It will be, in fact, because this state is stochastic and hence keeps on evolving indefinitely. 
%This is different from the NESS of the unconditional dynamics, which is the fixed point
%\begin{equation}\label{NESS}
%\rho_{X^*} = \mathcal{E}\big( \rho_{X^*} \big), 
%\end{equation} 
%of the map~\eqref{unconditional_map}.

%%%%%%%%%%%%%%%%%%%%%%%%%%%%%%
\subsection{Unconditional $\scnd$ law}\label{3c}

Next we turn to the thermodynamics. 
The $\scnd$ law of thermodynamics characterize  
the degree of irreversibility of a certain process and  
can be formulated in purely information-theoretic terms. 
This allows it to be extended beyond standard thermal environments, and also to avoid difficulties associated with the definition of heat and work, which can be quite problematic in the quantum regime~\cite{Landi2020a}. 

At each collision, the entropy of the system will change from $S(X_t)$ to $S(X_{t+1})$. 
This change, however, may be either positive or negative.
The goal of the $\scnd$ law is to identify a contribution to this change associated with the flow of entropy between system and ancilla, and another representing the entropy that was irreversibly produced in the process. 
The separation thus takes the form 
\begin{equation}\label{2nd}
\Delta \Sigma_t^u=  S(X_t) - S(X_{t-1}) +\Delta \Phi_t^u, 
\end{equation}
where $\Delta\Phi_t^u$ is the unconditional flow rate of entropy \emph{from} the system \emph{to} the ancilla in each collision, and $\Delta\Sigma_t^u$ is the unconditional rate of entropy produced in the process. 
The $\scnd$ law is  summarized by the statement that we should have $\Delta \Sigma_t^u\geq 0$. 
Eq.~\eqref{2nd} is merely a definition, however. 
The goal is precisely to determine the actual forms of $\Delta\Phi_t^u$ and $\Delta\Sigma_t^u$.

In standard thermal processes, this is usually accomplished by postulating that the entropy flow $\Delta \Phi_t^u$ should be linked with the heat flow $\dot{Q}_t$ entering the ancillae through Clausius' expression~\cite{Fermi1956} $\Delta\Phi_t^u = \beta \dot{Q}_t$, where $\beta$ is the inverse temperature of the thermal state the ancillae are in.
By fixing $\Delta \Phi_t^u$ we then also fix $\Delta \Sigma_t^u$. 
This, however, only holds for thermal ancillae, thus restricting the range of applicability of the formalism. 
 
Instead, we approach the problem using the framework developed in Ref.~\cite{Esposito2010a} (see also \cite{Manzano2017a,Strasberg2016}), which formulates the entropy production rate in information theoretic terms, as
\begin{equation}\label{sigma}
\Delta \Sigma_t^u= \mathcal{I}(X_t\!:\!Y_{t}') + D(Y_t'||Y_{t}) \geqslant 0, 
\end{equation}
where $\mathcal{I}(X_t\!:\!Y_{t}') = S(\rho_{X_{t}}) + S(\rho_{Y_{t}}')- S(\rho_{X_tY_t'})$ is the quantum mutual information between system and ancilla after Eq.~\eqref{global_map} and $D(Y_t'||Y_t) = D(\rho_{Y_t'} || \rho_{Y_t})$ is the relative entropy between the state of the ancilla before and after the collision.
The first term thus accounts for the correlations that built up between system and ancilla, while the second measures the amount by which the ancillae were pushed away from their initial states. 
Thus, from the perspective of the system, irreversibility stems from tracing over the ancillae after the interaction in such a way that all quantities related either to the local state of the ancilla, or to their global correlations, are irretrievable~\cite{Manzano2017a}. 
%This expression is non-negative by construction, $\Delta \Sigma_t^u\geqslant 0$.
%It also has an operational interpretation associated with which operations are allowed in the state after the collision, as discussed in Ref.~\cite{Manzano2017a}

As the global map in Eq.~\eqref{global_map} is unitary, and the system and ancillae are always uncorrelated before a collision, it follows that 
\begin{equation}
S(\rho_{X_tY_t'}) =S(\rho_{X_{t-1}Y_t})= S(\rho_{X_{t-1}}) + S(\rho_{Y_t}).
\end{equation}
Hence, the mutual information may also be written as
\begin{equation}\label{MI_sum_entropies}
\mathcal{I}(X_{t}\!:\!Y'_{t}) = S(X_t) + S(Y_{t}') - S(X_{t-1}) - S(Y_{t}). 
\end{equation}
Plugging this in Eq.~(\ref{sigma}) and comparing with Eq.~(\ref{2nd}) then allows us to identify the entropy flux as 
\begin{equation}
\label{phi}
\Delta\Phi_t^u = S(Y_{t}') - S(Y_{t}) + D(Y_{t}'||Y_{t})= \tr\Big\{ (\rho_{Y_t} - \rho_{Y_t'}) \ln \rho_{Y_t}\Big\}.
\end{equation}
The entropy flux is seen to depend solely on the degrees of freedom of the ancilla.
%Moreover, it also clearly describes a ``flow'', in the sense that it measures the difference in the ``thermodynamic potential'' $\langle \ln \rho_{Y_t}\rangle$, between the initial and final states of the ancilla. 
Although Eq.~(\ref{phi}) is  general and holds for arbitrary states of the ancillae, it reduces to $\beta \dot{Q}$, as in the Clausius expression, if $\rho_Y$ is thermal.

Another very important property of the entropy flux is additivity. 
What we call an ``ancilla'' may itself be a composed system consisting of multiple elementary units. 
In fact, as we will illustrate in Sec.~\ref{sec:qubits}, this can give rise to interesting situations. Suppose that $Y_t = (Y_{t1}, Y_{t2},\ldots, Y_{tN})$ and that the units are prepared in a globally product state 
$\rho_{Y_t} = \bigotimes^N_{j=1} \rho_{Y_{tj} }$. 
After colliding with the system, the state $\rho_{Y_t'}$ might no longer be uncorrelated, in general. 
Despite this, owing to the structure of Eq.~\eqref{phi}, we would have % the flux will be a sum of fluxes to each ancilla, 
\begin{equation}\label{phi_multiple}
\Delta \Phi_t^u = \sum\limits^N_{j=1} \Delta \Phi_{tj}^u =  \sum\limits^N_{j=1} \tr\Big\{(\rho_{Y_{tj}'} - \rho_{Y_{tj}}) \ln \rho_{Y_{tj}}\Big\},
\end{equation}
where $\rho_{Y_{tj}'}$ is the post-collision reduced state of the $j^\text{th}$ unit of the ancilla. 
This property is quite important, as it allows one to compute the flux associated to each dissipation channel acting on the system.

%%%%%%%%%%%%%%%%%%%%%%%%%%%%%%%%
\subsection{Conditional $\scnd$ law}\label{3d}

Eqs.~(\ref{2nd}), (\ref{sigma}) and (\ref{phi}) specify the thermodynamics of the unconditional trajectories $\rho_{X_t}$, when no information about the ancillae is recorded. 
We now ask the same question for the conditional trajectories $\rho_{X_t|\zeta_t}$. In this case, the relevant entropy is the quantum-classical conditional entropy $S(X_t|\zeta_t)$ in Eq.~\eqref{condS}. 
Thus, we search for a splitting analogous to Eq.~(\ref{2nd}), but of the form
\begin{equation}\label{2nd_cond}
\Delta\Sigma_t^c = S(X_{t}|\zeta_{t}) - S(X_{t-1}|\zeta_{t-1}) +\Delta \Phi_t^c,
\end{equation}
where $\Delta\Sigma_t^c$ and $\Delta \Phi_t^c$ are the conditional counterparts of the unconditional quantities used in Sec.~\ref{3c}.
The identification of suitable forms for such quantities is the scope of this Section. 

We adopt an approach similar to that used in Refs.~\cite{Breuer2003,Belenchia2019}, which consists in defining the conditional flux rate as the natural extension of Eq.~(\ref{phi}) to the conditional case.
That is, as $\Delta \Phi_t^c$ refers to a specific collision, it should depend only on quantities pertaining to the specific ancilla $Y_t$, thus being of the form 
\begin{equation}\label{phi_c}
\Delta\Phi_t^c = S(Y_t'|z_t) - S(Y_t) + \sum\limits_{z_t} P(z_t) D\big( \rho_{Y_t'|z_t} \big| \big| \rho_{Y_t}\big),
\end{equation}
%where the time subscripts have been omitted for clarity.
where $\rho_{Y_t'|z_t}=({M_{z_t} \rho_{Y_t'} M_{z_t}^\dagger})/{P(z_t)}$ is the final state of the ancilla given outcome $z_t$ and %, and is given by 
%\[
%\rho_{Y_t'|z_t} = \frac{M_{z_t} \rho_{Y_t'} M_{z_t}^\dagger}{P(z_t)},
%\]
$P(z_t) = \tr\big(M_{z_t} \rho_{Y_t'} M_{z_t}^\dagger)$ [cf. Eq.~\eqref{Pz}].
Moreover, $S(Y_t'|z_t)$ is defined similarly to Eq.~\eqref{condS}.
Note how the causal structure of the model implies that the flux should be conditioned only to outcome $z_t$, instead of the entire measurement record $\zeta_t$.

By defining the reconstructed state of the ancilla after the measurement $\tilde{\rho}_{Y_t'} = \sum_{z_t} P(z_t) \rho_{Y_t'|z_t} = \sum_{z_t} M_{z_t} \rho_{Y_t'} M_{z_t}^\dagger$, Eq.~\eqref{phi_c} can be recast into the form%, we can also absorb the sum over $\zeta_t$ and write $\Delta \Phi_t^c$ as
\begin{equation}\label{phi_c_tmp}
\Delta\Phi_t^c = \tr \Big\{ (\rho_{Y_t} - \tilde{\rho}_{Y_t'}) \ln \rho_{Y_t}\Big\},
\end{equation}
%where 
%\begin{equation}
%\tilde{\rho}_{Y_t'} = \sum\limits_{z_t} P(z_t) \rho_{Y_t'|z_t} = \sum\limits_{z_t} M_{z_t} \rho_{Y_t'} M_{z_t}^\dagger,
%\end{equation}
%is the reconstructed state of the ancilla after the measurement. 
which showcases the potential difference between conditional and unconditional fluxes. Depending on the measurement strategy $\{M_z\}$ being adopted, it is reasonable to expect that $\rho_{Y_t'}\neq\tilde{\rho}_{Y_t'}$, thus resulting in $\Delta\Phi_t^u\neq\Delta\Phi_t^c$. 
This reflects the potentially invasive nature of the measurements on the ancilla. 
%Different choices of $\{M_z\}$ can cause significant backaction, possibly making $\tilde{\rho}_{Y_t'}$ significantly different from $\rho_{Y_t'}$. 
However, it should be noted that this is an extrinsic effect, related to the specific choice of measurement by the observer, and fully unrelated to the thermodynamics of the system-ancilla interactions. %, but with an extraneous hypothesis on how the experiment is carried out. 

We will henceforth assume that the  measurement strategy is such that %It is also not difficult to implement strategies which generate no additional entropy flux. 
%We do not need to have  $\tilde{\rho}_{Y_t'} = \rho_{Y_t'}$;all we need is that 
\begin{equation}\label{measurement_condition}
\tr\{\tilde{\rho}_{Y_t'} \ln \rho_{Y_t}\} = \tr\{\rho_{Y_t'} \ln \rho_{Y_t}\}.
\end{equation}
That is, it that does not change the population of $Y_t'$ in the eigenbasis of the \emph{original} state $\rho_{Y_t}$. 
This can be accomplished, for instance, by measuring in the same basis into which the state of the ancillae is prepared. % ensure that this is always satisfied. 
%In the case when Eq.~\eqref{measurement_condition} is satisfied, 
We can then reach the important conclusion that 
\begin{equation}\label{phi_c_uc}
\Delta\Phi_t^c =\Delta \Phi_t^u.
\end{equation} 
%that is, the entropy flux in the conditional and unconditional dynamics are the same. 
This result is intuitive: Conditioning on the outcome is a subjective matter, related to whether or not we read out the outcomes of the experiment. 
It should therefore have no effect on how much entropy flows to the ancillae.
Similar ideas were also used in many contexts~\cite{Breuer2003,Sagawa2012,Funo2013,Strasberg2019a}. 
However, these studies were concerned with the heat flux, which coincides with the entropy flux for thermal baths. Here we show that this is a general property, valid for any bath, provided we restrict to the special class of measurements characterized by Eq.~\eqref{measurement_condition}. 
%In 
%Ref.~\cite{Belenchia2019} the same result was proved in the particular case of Gaussian systems and measurements.%, a case  The present general result extend such derivation beyond the Gaussian case.
%While we will not consider this case in the present work, it should be noticed that this mismatch is a purely quantum effect since it is not present for conditional incoherent dynamics (see App....)

Under these conditions, %, plugging Eq.~(\ref{phi_c_uc}) into Eq.~(\ref{2nd_cond})  yields the conditional entropy production 
%\begin{equation}\label{sigma_c}
%\Delta\Sigma_t^c = S(X_t|\zeta_t) - S(X_{t-1}|\zeta_{t-1}) + \Delta\Phi_t. 
%\end{equation}
comparing Eqs.~\eqref{2nd_cond} and ~(\ref{2nd}), and reminding of the information rate in Eq.~\eqref{Delta_I}, we find  %shows that the two quantities are actually related by 
\begin{equation}\label{two_sigmas}
\Delta\Sigma_t^c = \Delta \Sigma_t^u- \Delta I_t.
\end{equation}
%where $\Delta I_t$ is the Holevo information rate~\eqref{Delta_I}. 
This is a key result of our framework: It shows how the act of conditioning the dynamics on the measurement outcome changes the entropy production by a quantity associated with the change in the Holevo information. 
Hence, it serves as a bridge between the information rates and thermodynamics. 
In particular, in an ISS, $\Delta I_\text{ISS} = 0$ and so 
%In an informational steady-state (ISS), as we have seen, we have $\Delta I_t = 0$ but $G_\text{ISS} = L_\text{ISS} \neq 0$ [Eq.~\eqref{ISS}]. 
%Hence, in the ISS we must have
%\begin{equation}
$\Delta \Sigma_\text{ISS}^c = \Delta \Sigma_\text{ISS}^u$, 
%\end{equation}
although $\rho_{X_t}$ and $\rho_{X_t|\zeta_t}$ are in general different. 

%%%%%%%%%%%%%%%%%%%%%%%%%%%%%%%%
\subsection{Properties of the conditional entropy production}

%Concerning the unconditional dynamics, the system is said to be in a non-equilibrium steady-state (NESS) if $\Delta \Sigma^u \neq 0$, or in an equilibrium state if $\Delta \Sigma^u = 0$. 
%ISSs and NESSs are unrelated, since the former depends on the measurements and the latter does not. 
%But, as this result shows, if the unconditional dynamics is in equilibrium, we must have $\Delta \Sigma_\text{ISS}^c = 0$, even if the system is in an ISS.
%An example of this  was studied experimentally in~Ref.~\cite{Rossi2020} and will be discussed further  in Sec.~\ref{sec:gaussian}.

%A non-zero entropy production rate is the signature of a non-equilibrium steady-state (NESSs). 
%Hence, this shows that ISSs are a subclass of NESSs. 
%But, more importantly, it shows that ISSs can only exist if the unconditional system is in a NESS. 
%That is, if $\Delta \Sigma^u \neq 0$. 
%If the unconditional system eventually thermalizes, then so will the 

We now move on to discuss the main properties of the conditional entropy production.
The quantities $\Delta \Sigma_t^u$ and $\Delta \Sigma_t^c$ refer to the incremental entropy production in a single collision. 
Conversely, it is also of interest to analyze the integrated entropy production 
%we now show that if one considers instead the integrated entropy productions 
\begin{equation}
\Sigma_t^\alpha = \sum\limits_{\tau=1}^t \Delta \Sigma_\tau^\alpha,\qquad \alpha = u,c. 
\end{equation}
Since $\Delta I_t$ in Eq.~\eqref{Delta_I}  is an exact differential, when we sum Eq.~\eqref{two_sigmas} up to time $t$, the terms in $\Delta I_\tau$  successively cancel, leaving only
\begin{equation}\label{integrated_sigmas}
\Sigma_t^c = \Sigma_t^u - I(X_t \! : \! \zeta_t).  
\end{equation}
The integrated entropy production up to time $t$ therefore depends only on the net information $I(X_t\! : \! \zeta_t)$. 
Since $I(X_t\! : \! \zeta_t) \geqslant 0$, it then follows that
\begin{equation}\label{integrated_sigma_bound_cu}
%\Sigma_t^c \leqslant \Sigma_t^u. 
\Sigma_t^u \geqslant \Sigma_t^c. 
\end{equation}
Therefore, \emph{conditioning makes the process more reversible}. 
This happens because we only carry out measurements in the environment, so that there is never a direct backaction in the system. 
A stronger bound can also be obtained by using the fact that $L_t \geqslant 0$, which then leads to 
\begin{equation}
\Sigma_u - \Sigma_c \geqslant \sum\limits_{\tau=1}^t G_\tau. 
\end{equation}
The reduction in entropy production is thus \emph{at least} the total information gain. 

Returning now to the entropy production rate in each collision, in Appendix~\ref{app:proof_2ndlaw_conditional} we provide a proof of the following relation  
\begin{equation}\label{delta_sigma_c_inequality}
\Delta \Sigma_t^c \geqslant D(Y_{t}' || Y_{t})  + I(Y_{t}' \! : \! \zeta_{t}) \geqslant 0,
\end{equation}
where $D(Y_{t}' || Y_{t})$, is the backaction caused in the ancillary state due to its collision with the system, while $I(Y_{t}' \! : \! \zeta_{t-1})$ quantifies the amount of information gained about the ancilla through the measurement strategy. 
This is one of the overarching conclusions of our work, bearing remarkable consequences. 
On the one hand, it proves that the $\scnd$ law continues to be satisfied in the conditional case. 
On the other hand, it provides a non-trivial lower bound to the conditional entropy production rate in terms of the changes that take place in the \emph{ancillae} only. 
It should also be noted that, the first inequality in Eq.~\eqref{delta_sigma_c_inequality} is saturated by processes where the measurement extracts all the information available.

%%%%%%%%%%%%%%%%%%%%%%%%%%%%%%%%%%%%%%%%%%
%
%
\section{\label{sec:qubits}Simple qubit models}
%
%
%%%%%%%%%%%%%%%%%%%%%%%%%%%%%%%%%%%%%%%%%%

We now apply the ideas of the previous sections to simple models of $\text{CM}^2$s, aimed at illustrating their overarching features while keeping the level of technical details to a minimum, so as to emphasize the physical implications of the framework illustrated so far. 

We will focus on the case in which both the system and the elementary units of the ancilla are qubits. Despite their simplicity, such situations have far-reaching applications. 
For instance, in  Ref.~\cite{Gross2017a} it was shown how quantum optical stochastic master equations naturally emerge from modeling opticals baths in terms of effective qubits in a collisional model. 
%Via this ``qubit-ization'' approach, one can study several light-matter scenarios. 
Moreover, suitably chosen measurement stategies $\{M_z\}$ implemented on qubits allow also to simulate widely used measurement schemes, such as photo-detection, homodyne and heterodyne measurements. Finally, by tuning the initial state of the qubits, one can also simulate out-of-equilibrium environments, such as squeezed baths. In Ref.~\cite{tofollow}, we complement the study reported here by addressing explicitly the case of continuous-variable systems.

%We provide a publicly available library for simulating qubit-based.
Recall that a $\text{CM}^2$ is completely specified by setting  $\{ \rho_Y, U, M_z\}$.
The unconditional dynamics is governed by the map $\mathcal{E}$ defined in Eq.~\eqref{unconditional_map}, which can be simulated directly with very low computational cost. 
The conditional dynamics, on the other hand, is governed by the map $\mathcal{E}_z$ in Eqs.~\eqref{conditional_map_def} and~\eqref{conditional_map}, which we simulate using stochastic trajectories.

%%%%%%%%%%%%%%%%%%%%%%%%%%%%%%%%%%%%%%%%%%
\subsection{\label{ssec:single_qubit}Single-qubit ancilla}

\begin{figure*}
\includegraphics[width=0.9\textwidth]{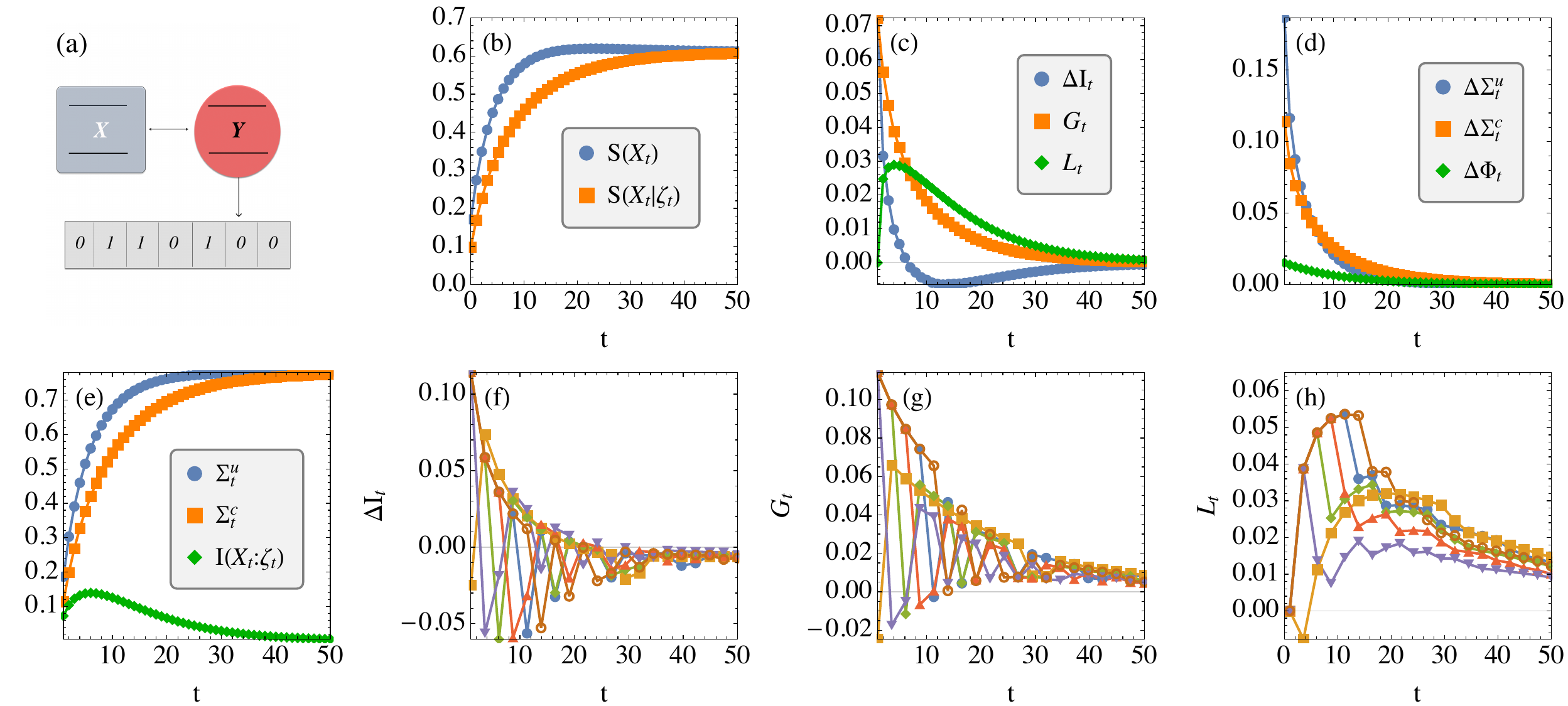}
\caption{
(a) Dynamics of a $\text{CM}^2$ under a quantum homogenization process where both system and ancilla are qubits.
(b) Unconditional and conditional entropies. 
(c) Information rate $\Delta I_t$ [Eq.~\eqref{Delta_I}], and its splitting into a gain and loss term [Eq.~\eqref{Delta_I_splitting}].
(d) Unconditional and conditional entropy production rates, $\Delta \Sigma_t^u$ and $\Delta \Sigma_t^c$, as well as the entropy flux $\Delta \Phi_t$. 
(e) Integrated unconditional and conditional entropy productions, and net Holevo information $I(X_t\! : \! \zeta_t)$ [c.f. Eq.~\eqref{integrated_sigmas}].
(f), (g), (h) Sample stochastic trajectories of $\Delta I_t$, $G_t$ and $L_t$. 
We have taken $f=g=0.3$ (the results do not depend qualitatively on such choices) while details on how we chose $\rho_Y$, $\rho_{X_0}$, $U$ and $\{M_z\}$ are explained in the main text.
}
\label{fig:single_qubit}
\end{figure*}

We begin by studying the  case where the system interacts with single-qubit ancillae
prepared  in the thermal state 
$\rho_Y = f|0\rangle\langle 0 |_Y + (1-f) |1\rangle\langle 1|_Y$, where 
$f\in[0,1]$ and  $\ket{0}, \ket{1}$ is the computational basis --- i.e., the eigenstates of the Pauli-$z$ operator $\sigma^z_Y=|1\rangle\langle 1 |_Y-|0\rangle\langle 0 |_Y$. 
The collisions are modeled by a partial SWAP gate $U = e^{- i g(\sigma_X^+ \sigma_Y^- + \sigma_X^- \sigma_Y^+)}$, where $\sigma_{\alpha}^+=\left(\sigma_{\alpha}^-\right)^\dagger=\ket{1}\bra{0}_\alpha$ is the Pauli raising operator ($\alpha=X,Y$).
Finally, we assume that the ancillae are measured in the computational basis, so that 
$M_0 = |0\rangle\langle 0 |_Y$ and $M_1 = |1\rangle\langle 1|_Y$. 
For concreteness, we take the initial state of the system to be 
$\rho_{X_0} = |x_+\rangle\langle x_+|_X$, where $\sigma^x_X|x_+\rangle_X= |x_+\rangle_X$. 
%We also fix the parameters of the model at $f = g = 0.3$. 

The evolution of the relevant information and thermodynamic quantities of the problem, for a specific choice of $f$ and $g$, is presented in Fig.~\ref{fig:single_qubit}.
Panel (b) shows 
how conditioning always reduces our ignorance about the system,  by demonstrating that $S(X_t|\zeta_t) \leqslant S(X_t)$ at all times.
As the model being considered implement a homogenization process~\cite{Scarani2002,Ziman2002}, the steady state $\rho_{X_\infty}$  coincides with the initial state of the ancilla, $\rho_{X_\infty} = \rho_Y$. 
This causes $U(\rho_{X_\infty} \otimes \rho_Y)U^\dagger = \rho_{X_\infty} \otimes \rho_Y$, so  no information can be acquired anymore.
The final state is thus an equilibrium state, not an ISS. 
The information rate, gain and loss are shown in Fig.~\ref{fig:single_qubit}(c). 
Initially the gain is very large, as the state of the system is significantly different from the thermal steady state and each measurement results in a significant acquisition of information. In turn, this results in $\Delta I_t > 0$. As the system evolves towards $\rho_{X_\infty}$, the detrimental effect of homogenization starts prevailing over the information gain, causing an inversion in the sign of $\Delta I_t$. The long-time limit is associated with $(\Delta I_\infty, G_\infty, L_\infty)\to 0$ and no ISS emerges.

A comparison between the conditional and unconditional entropy production is shown in Fig.~\ref{fig:single_qubit}(d), which also reports on the entropy flux. 
The rates $\Delta \Sigma_t^c$ and $\Delta \Sigma_t^u$ are both non-negative, but are not necessarily ordered. 
 This happens because, in individual collisions, conditioning may not make the process more reversible. An ordering is instead enforced when looking at integrated quantities: Conditioning always reduces the entropy production [cf. Eq.~\eqref{integrated_sigmas}], as shown in Fig.~\ref{fig:single_qubit}(e). %, where we also compare them with the net information $I(X_t \! : \! \zeta_t)$. 

For completeness, we also show in Figs.~\ref{fig:single_qubit}(f), (g), (h) the behavior of $\Delta I_t$, $G_t$ and $L_t$ along six randomly sampled trajectories $\zeta_t$. 
Typical stochastic fluctuations are observed, showing that in a single stochastic run, the net gain and loss can differ substantially (the curves in Fig.~\ref{fig:single_qubit}(b)-(e) were produced by averaging over 2000 such trajectories).

%%%%%%%%%%%%%%%%%%%%%%%%%%%%%%%%%%%%%%%%%%
\subsection{\label{ssec:two_qubit_ancilla}Two-qubit ancilla}

We now move on to consider a case %where the system is still a single qubit, but the ancilla is composed of two qubits, prepared in different states, only one of which is actually measured. As we show, these models serve as building blocks for constructing 
allowing the emergence of ISSs, opening up many interesting possibilities.
\begin{figure*}
\includegraphics[width=0.9\textwidth]{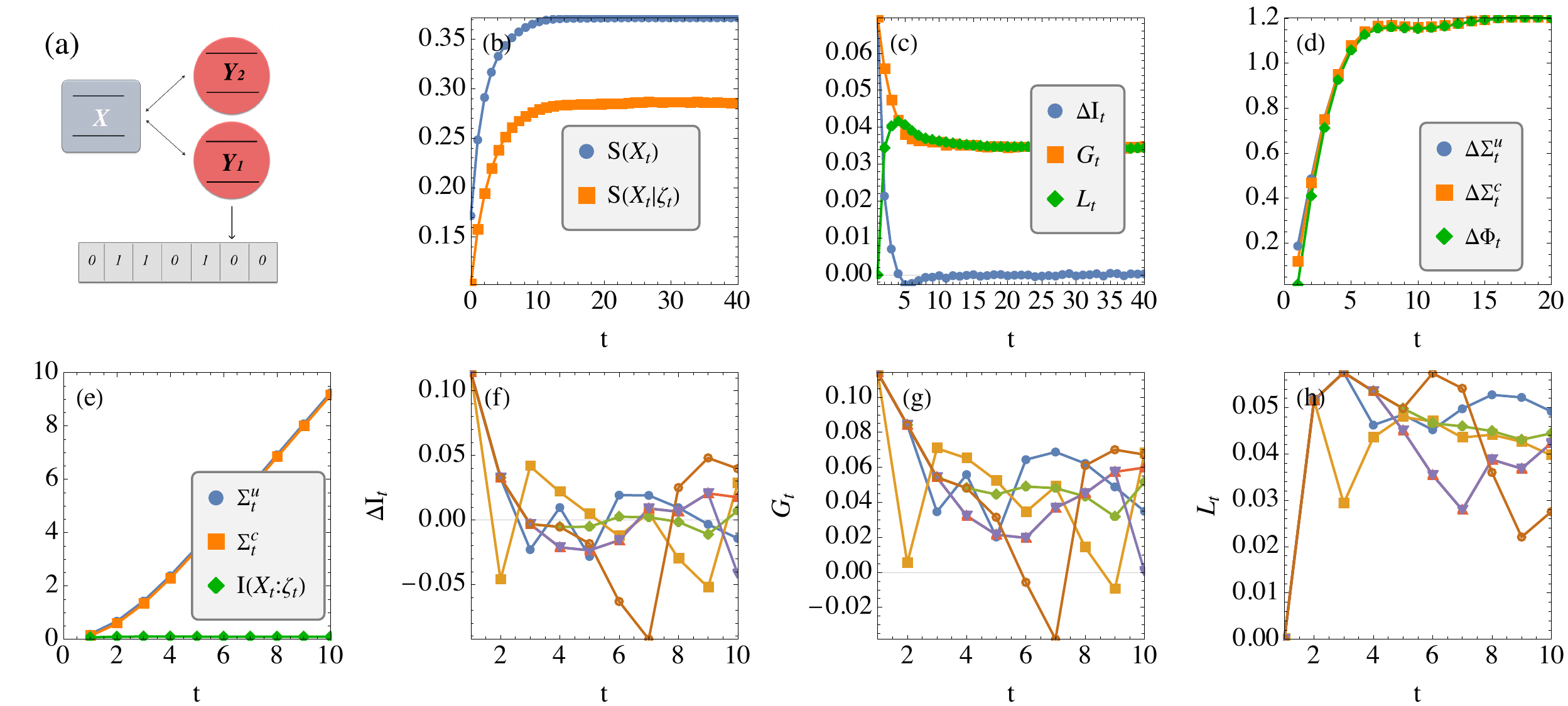}
\caption{
Same as Fig.~\eqref{fig:single_qubit}, but for two-qubit ancillae, prepared in 
$\rho_{Y^1} = f |0\rangle\langle 0 | + (1-f) |1\rangle\langle 1|$ and 
$\rho_{Y_2} = |x_+\rangle\langle x_+|$. The qubits interact sequentially with the system via partial SWAPs and only ancilla $Y_1$ is measured. 
In contrast to Fig.~\eqref{fig:single_qubit},  this model has a non-trivial ISS ($G = L \neq 0$). We have taken, for concreteness, $f = g_1 = 0.3$ and $g_2 = 0.1$. 
}
\label{fig:two_qubits_ex1}
\end{figure*}
The ancillae do not have to be just a single qubit, but can have arbitrary internal structure. 
Moreover, within a single collision, the system does not have to interact with all elementary units simultaneously, but may do so sequentially. 
We illustrate this by considering the case where each ancilla is actually 2 qubits, $Y_t = (Y_{t1},Y_{t2})$, which interact sequentially with the system (cf. Fig.~\ref{fig:two_qubits_ex1}).
The unitary $U_t$ between $X$ and $Y_t$ will then have the form 
\begin{equation}\label{qubits_composition_unitaries}
U_t = U_{XY_{t2}} U_{XY_{t1}}, 
\end{equation}
where $U_{XY_{tj}}$ has support only over the Hilbert space of $X$ and the unit $Y_{tj}$. 
As discussed in Ref.~\cite{Rodrigues2019,Landi2020a}, if the ancillae are prepared in different states, the system will not be able to equilibrate with either,  but will instead keep on bouncing back and forth indefinitely. Hence, it will reach a NESS.
Moreover, if at least one of the ancillae are measured, the conditional state may embody an ISS.

To illustrate this, we assume the first unit to be prepared in a thermal state such as the one considered  in Sec.~\ref{ssec:single_qubit}, while the second unit is in $|x_+\rangle$.
The unitaries in Eq.~\eqref{qubits_composition_unitaries} are chosen, as before, to be partial SWAPs with strengths $g_1$ and $g_2$.
Finally, we choose to measure only the first unit which, by being prepared in a thermal state, acts as a classical probe. On the other hand, by being endowed with quantum coherence, the second unit  represents a ``resourceful state.''
%~\footnote{It is interesting to note that, for composite ancillary systems, measuring only one of the subsystems still guarantees the validity of Eq.~\eqref{phi_c_uc}, as far as the initial state of the ancilla is separable in the partition $Y_{1;t}\otimes Y_{2,3,...N;t}$  (with $\{Y_{i;t}\}$ the different subsystems of the single ancilla $Y_t$) and we measure in the basis of $\rho_{Y_{1;t}}$. This means that, in order to satisfy Eq.~\eqref{phi_c_uc} in this particular case, no entanglement should be present in the initial state of the ancilla between the subsystem which is monitored and the rest of the subsystems.   
%}. 

In Fig.~\ref{fig:two_qubits_ex1} we report the results of an analysis similar to the one that we have performed for the previous example, for direct comparison. 
The results are strikingly different as, in particular, the system now allows for an ISS. 
This is visible in Fig.~\ref{fig:two_qubits_ex1}(c) from the fact that $G = L \neq 0$ when $t\to \infty$ with the thermodynamic quantities in Fig.~\ref{fig:two_qubits_ex1}(d) also converging to non-zero long-time values. 
A marked difference with the case of no ISS is also seen in the behavior of the integrated entropy production in Fig.~\ref{fig:two_qubits_ex1}(e):
As the rates now remain non-zero, the integrated quantities diverge in the long-time limit. 

\begin{figure*}
\includegraphics[width=0.9\textwidth]{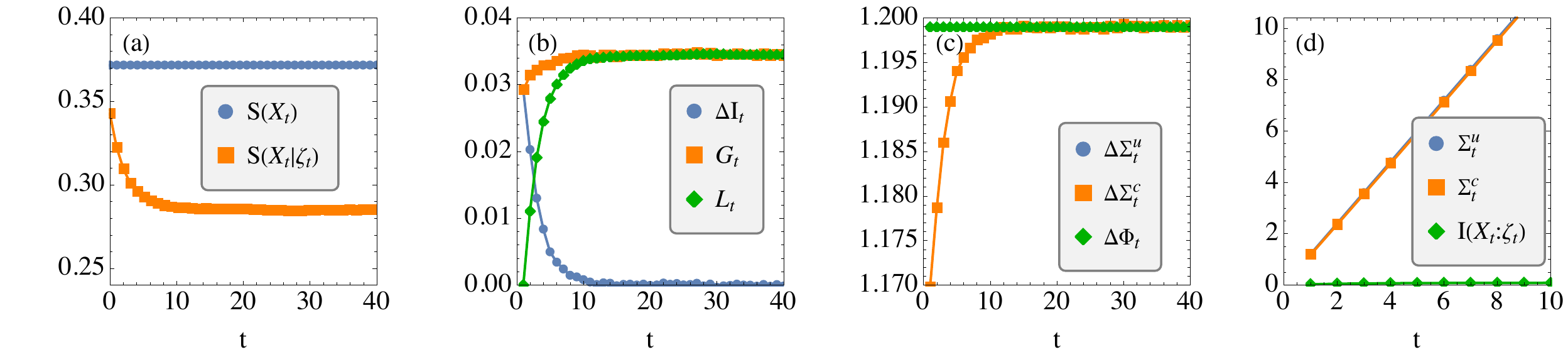}
\caption{
Similar to the two-qubit scenario of Figs.~\ref{fig:two_qubits_ex1}(b)-(e), but with the initial state $\rho_{X_0}$ chosen as the fixed point $\rho_{X^*}$ of the unconditional dynamics. 
}
\label{fig:two_qubits_ex2}
\end{figure*}

We can also perform another experiment that beautifully illustrates the essence of an ISS. 
While the initial state used in Fig.~\ref{fig:two_qubits_ex1} was arbitrarily chosen, we could take it to be the steady-state of the unconditional dynamics. 
%That is, the state which satisfies 
%\begin{equation}\label{fixed_point_uncond}
%\rho_{X^*} = \mathcal{E}(\rho_{X^*}), 
%\end{equation}
%where $\mathcal{E}$ is given in Eq.~\eqref{unconditional_map}. 
The idea is that we first allow the system to unconditionally relax by letting it undergo a large number of collisions, and only then we start measuring. 
%What will happen is that, unconditionally, the state will no longer change since it has already reached a fixed point. That is, $\rho_{X_t} = \rho_{X^*}$.  
%But \emph{conditionally}, it will start to evolve. 
Due to the effect of the measurements, the conditional state $\rho_{X_t|\zeta_t}$ will start to differ from unconditional steady-state (while the unconditional dynamics remains fixed).

The results are shown in Fig.~\ref{fig:two_qubits_ex2}.
Panel (a), in particular, neatly illustrates how the unconditional entropy does not change in time, while the measurements performed in the conditional strategy   reduce the entropy of the state of the system, which is effectively driven to a state with a larger purity. {This is the essence of an ISS.} 
%This colder conditional state $\rho_{X_\infty|\zeta_\infty}$ only continue to exist if we keep measuring it. 
%As soon as we turn the measurement off, the system would heat up again back to $S(X^*)$.
%This idea is corroborated by the gain and loss rates in Fig.~\ref{fig:two_qubits_ex2}(b) and the thermodynamic quantities in 
%Fig.~\ref{fig:two_qubits_ex2}(c) and (d).  

%%%%%%%%%%%%%%%%%%%%%%%%%%%%%%%%%%%%%%%%%%
\subsection{Time series in the single-shot scenario}

\begin{figure*}
\includegraphics[width=\textwidth]{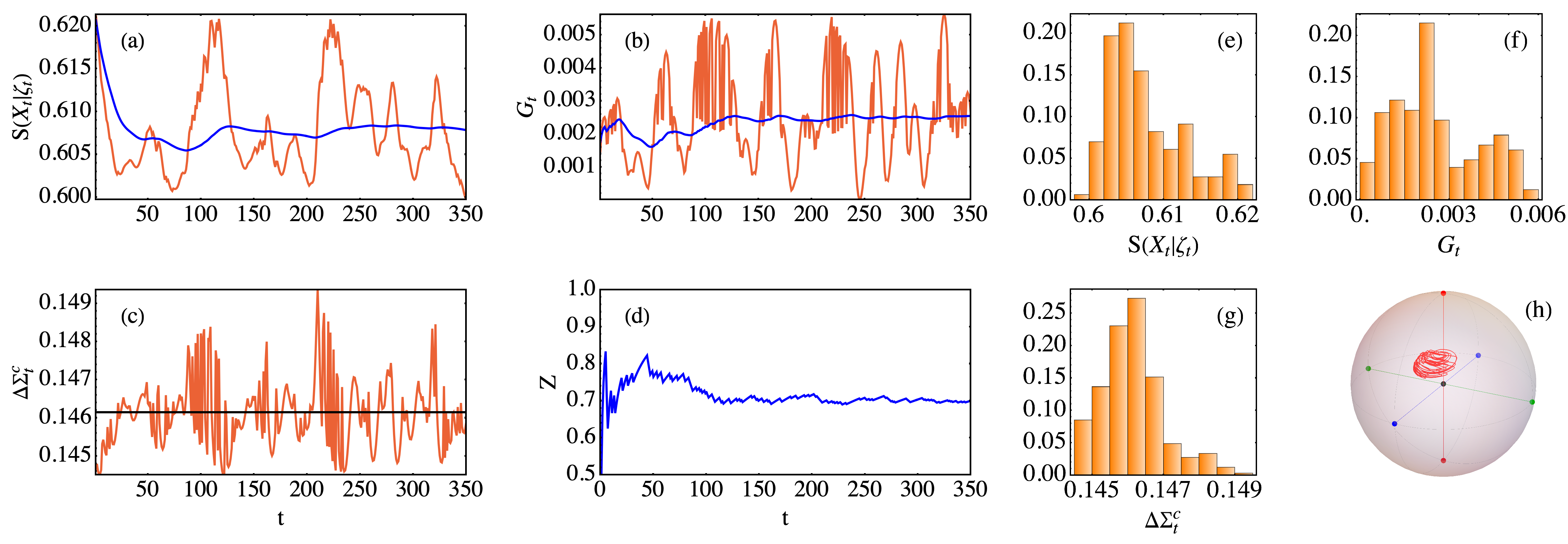}
\caption{
Thermodynamics and information in the single-shot scenario. 
The configuration is the same as Fig.~\ref{fig:two_qubits_ex2}, but everything now refers to a single stochastic realization of the experiment.
The red curves depict (a) $S(X_t|\zeta_t)$, (b) $G_t$ and (c) $\Delta \Sigma_t^c$ for that single realization. 
The blue curves, on the other hand, represent the accumulate average; that is, the average of the given quantity up to that time. 
Image (d), in particular, shows the accumulated average for the outcomes $Z_t = (\sum_{j=1}^t z_t)/t$, where the outcomes $z_t$ are either $0$ or $1$ (not shown for visibility). 
The black line in image (c) is the unconditional entropy production rate $\Delta \Sigma_t^u$, which serves as a baseline for $\Delta \Sigma_t^c$. 
Images (e)-(g) are the histograms obtained from the data in (a)-(d), discarding the first 20 points (to eliminate transients).
(h) Stochastic trajectory in Bloch's sphere. 
}
\label{fig:time_series}
\end{figure*}

The quantities in Fig.~\ref{fig:single_qubit}-\ref{fig:two_qubits_ex2} were obtained by repeating the experiment multiple times, always starting from the same state and evolving in the exact same way. 
We now contrast this with the single-shot scenario. 
That is, when we have access only to a single stochastic realization of the experiment. 
We focus on the two-qubit model where the system starts in the steady-state of the unconditional dynamics, as in Fig.~\ref{fig:two_qubits_ex2}. 
The dynamics of $S(X_t|\zeta_t)$, $G_t$ and $\Delta \Sigma_t^c$ along a single trajectory is shown in Fig.~\ref{fig:time_series}. 
As one might expect, these quantities fluctuate significantly.
%The conditional entropy, in particular, starts in $S(X^*)$ and, for short times, decays similar to the average  in Fig.~\ref{fig:two_qubits_ex2}(a). 
%But after a certain time, revivals  occur, which can cause $S(X_t|\zeta_t)$ to fluctuate significantly. 
%Similar conclusions can be drawn from $G_t$ and $\Delta \Sigma_t^c$. 

Fig.~\ref{fig:time_series} also shows the behavior of accumulated averages, up to a certain time, showing that  both the entropy and gain rate % That is, the average of that stochastic realization up to that given time. 
%Thus, one sees for instance that the accumulated average of $S(X_t|\zeta_t)$ 
tend to converge precisely to the ISS value in Fig.~\ref{fig:two_qubits_ex2}. 
%And similarly for $G_t$. 
In a classical context, processes satisfying this property are called stationary ergodic~\cite{Peebles1993}. 
In Fig.~\ref{fig:time_series}(d) we plot the integrated average of the actual outcomes, $Z_t = (\sum_{j=1}^t z_t)/t$, the actual outcomes being binary.
%We don't plot the actual outcomes $z_t$, since they are  either 0 or 1, and so are not very easy to visualize.
Such integrated average outcome shows that in the ISS 70\% of the clicks are associated with $M_1$ and the remaining $30\%$ with $M_0$.

Finally, the single-shot data in Fig.~\ref{fig:time_series}(a)-(d) can also be used to construct a histogram of the most relevant quantities, as illustrated in panels (e)-(h). These histograms shed light on the magnitude of the fluctuations of the relevant quantities. For instance, $\Delta \Sigma_t^c$ fluctuates very little, while the information gain $G_t$ fluctuates dramatically. 

%%%%%%%%%%%%%%%%%%%%%%%%%%%%%%%%
%
%
%\section{\label{sec:comparison}Comparison with the extant literature}
%
%
%%%%%%%%%%%%%%%%%%%%%%%%%%%%%%%%

%This will be even more apparent from the next section in which we consider continuous variable models with the ancilla in pure non-equilibrium states for which a notion of temperature is not defined. More importantly, in our framework, the monitoring of the system happens by measuring (part of) the environment represented by the ancillae. It is due to this fact that we can characterize the thermodynamic and informational effect of repeated weak measurements by comparing with the unmeasured system. And this comparison is at the basis of our definition of ISS and the differential informational gain. Thus, the main difference with Refs~\cite{Strasberg2019c,Strasberg2020} is in the link between informational and thermodynamic quantities that we explore in our work in contrast to the citied references mostly focused on the energetic and other outstanding problems in quantum thermodynamics. 

%%%%%%%%%%%%%%%%%%%%%%%%%%%%%%%%
%
%
\section{\label{sec:conc} Conclusions}
%
%
%%%%%%%%%%%%%%%%%%%%%%%%%%%%%%%%

We have investigated the interplay between information and thermodynamics in continuously measured system by way of a collisional model construct. In particular, we were able to formulate the entropy production and flux rate --- two pivotal quantities in (quantum) thermodynamics --- from a purely informational point of view and accounting for repeated indirect measurements of the system of interest. These results offer a clear way to point-out and characterise the effect of quantum measurements on the thermodynamics of open quantum system. %Moreover, they generalized recent theoretical and experimental results~\cite{Belenchia2019, Rossi2020} beyond Gaussian systems and dynamics. 

We model the indirect measurement of the system via a collisional model where (a part of) the environment with which the system interact is monitored. This allows us to compare the entropy production with the case in which the environment is not measured and the evolution of the system is thus unconditioned. In turn, this comparison leads directly to a tightened second law for monitored systems with a very clear separation between entropic contributions coming from the dissipative interaction with the environment and the ones coming from the information gained during the monitoring. 
This allows us to introduce the concept of information gain rate and loss rates, and informational steady-states. The latter are particularly interesting since they represent cases where a delicate balance is established between the information that gets lost into the environment and the one that is extracted by measuring. 

%It is thus worth it to clarify the similarities with, as well as the crucial novelties of, our framework.% differs from those works and what are the common trends.%original contributions we introduce. 
%In a nutshell, the main features of our framework are: 
%\begin{itemize}
%\item Centered around the Holevo $\chi$ quantity, one of the most widely used quantities in quantum information theory. 
%\item Separates the information rate into a gain and loss rate, with clear interpretations. 
%\item Designed specifically describing for ISSs. 
%\item Extends thermodynamics beyond thermal baths, to generic non-equilibrium environments. 
%\item Flexible collisional model construct, which does not rely on specific master equations. 
%\end{itemize}
The interplay between information and the $\scnd$ law has been the subject of several works over the last decade. 
Stroboscopic dynamics, such as the one considered in Sec.~\ref{sec:setup}, have been studied in the  classical context of Hidden Markov models~\cite{Darwiche2009,Neapolitan2003,Ito2013}. 
A classical framework, where quantum measurements are mimicked by generic interventions, was put forth in~\cite{Strasberg2019a}, and resembles the classical version of our $\text{CM}^2$'s, developed in Appendix~\ref{app:incoherent_models}.
%They can also be viewed as a generalization of  the formalism in~\cite{Ito2013}.
%To some extent, our work can be viewed as the quantum generalization of the seminal work~\cite{Ito2013}. 
In the quantum context, the conditional dynamics analyzed here are a particular case of process tensors~\cite{Chiribella2008a,Pollock2018b,Pollock2018} whose thermodynamics has been recently considered in~\cite{Strasberg2019c,Strasberg2020}. Unlike our framework, however, these studies assume the system is always connected to a standard thermal bath, while the ancillae play only the role of memory agents. For this reason, their definition of entropy production is based on a Clausius-like inequality and is therefore different from ours. Furthermore, we have opted to focus  on informational aspects of thermodynamics, neglecting entirely the \emph{energetics} of the problem. Detailed accounts of the latter can be found in Ref.~\cite{Alonso2016,Strasberg2019c,Strasberg2020}.
%\\
%Crucially, the  comparison between unconditional and conditional entropy production as well as the notion of an ISS and the tools to characterize it are missing in these works. 
%\\

%Finally, Ref.~\cite{Funo2013} --- recently assessed experimentally in Ref.~\cite{Naghiloo2020} --- also describes quantum mechanically how the acquisition of information through a memory device affects the entropy production. Ref.~\cite{Funo2013} focuses, however, on a single measurement and does not account for continuously monitored systems and their ramifications, such as ISSs. Furthermore, in Ref.~\cite{Funo2013} information is acquired through a memory which interacts with the system. The role of the memory is thus active and always deleterious, since it necessarily causes a certain amount of backaction. In our formalism, since we assume generically out of equilibrium ancillae, the roles of the memory and the ancillae are  quite similar but they interact with the system whether or not we measure them. The measurement on the ancillae is therefore passive: it can only offer some bonus information about the system. 

Ref.~\cite{Funo2013} put forth a framework (recently assessed experimentally in Ref.~\cite{Naghiloo2020}) where the ancillae play the role of active memories.
This means their  effect is always deleterious to the system.
As a consequence, instead of using the Holevo quantity~\eqref{holevo} to quantify information, they use the Groenewold-Ozawa quantum-classical information~\cite{Groenewold1971,Ozawa1986} $I_{GO} = S(X) - S(X'|z)$. The two quantities are related by $I(X'\! : \! z) = I_{GO} - \Delta S_X$, where $\Delta S_X = S(X') - S(X)$.
Depending on the type of collision, $\Delta S_X$ may have any sign, so $I_{GO}$ is not necessarily non-negative.

%This difference is also reflected in how our framework and Ref.~\cite{Funo2013} quantify information. Whether we compare the entropy of the system with and without conditioning on the outcomes through the Holevo quantity~\eqref{holevo}, Ref.~\cite{Funo2013} compares the conditioned state with the \emph{initial} state -- instead of the final unconditioned one -- 
%Our framework is centered on the Holevo quantity~\eqref{holevo}, which compares the entropy of the system with and without conditioning on the outcomes.
%For a single collisions $X \to (X',z)$  it reads $I(X'\! : \! z) = S(X') - S(X'|z)$. 
%Ref.~\cite{Funo2013}, on the other hand, quantifies information through the so-called Groenewold-Ozawa quantum-classical information~\cite{Groenewold1971,Ozawa1986}, which in our notation reads
%\[
%I_{GO} = S(X) - S(X'|z). 
%\]
%That is, it compares $X'|z$ with the \emph{initial} state $X$, instead of the final state $X'$. 
%It therefore quantifies not only the information acquired, but also the backaction that caused $X$ to change to $X'$. 
%The two quantities are related by $I(X'\! : \! z) = I_{GO} - \Delta S_X$, where $\Delta S_X = S(X') - S(X)$.
%Depending on the type of collision, $\Delta S_X$ may have any sign. 
%For this reason $I_{GO}$ is not necessarily non-negative, while the Holevo quantity is. 

The formalism developed in this work is widely applicable,  as exemplified by the case studies  we have considered (see also~Ref.~\cite{tofollow}).
This makes it a valuable tool in the thermodynamic assessment of a broad variety of quantum-coherent experiments. 
The scenario we considered also fits perfectly with the characterization of  emergent quantum applications, such as quantum computing devices~\cite{gardas2018quantum,buffoni2020thermodynamics,cimini2020experimental}.
Being able to characterize irreversibility in these devices should thus offer a significant advantage in the design and engineering of future devices.

% @article{gardas2018quantum,
%  title={Quantum fluctuation theorem for error diagnostics in quantum annealers},
%  author={Gardas, Bart{\l}omiej and Deffner, Sebastian},
%  journal={Scientific reports},
%  volume={8},
%  number={1},
%  pages={1--8},
%  year={2018},
%  publisher={Nature Publishing Group}
%}
%@article{buffoni2020thermodynamics,
%  title={Thermodynamics of a quantum annealer},
%  author={Buffoni, Lorenzo and Campisi, Michele},
%  journal={Quantum Science and Technology},
%  volume={5},
%  number={3},
%  pages={035013},
%  year={2020},
%  publisher={IOP Publishing}
%}
%@article{cimini2020experimental,
%  title={Experimental characterization of the energetics of quantum logic gates},
%  author={Cimini, Valeria and Gherardini, Stefano and Barbieri, Marco and Gianani, Ilaria and Sbroscia, Marco and Buffoni, Lorenzo and Paternostro, Mauro and Caruso, Filippo},
%  journal={npj Quantum Information},
%  volume={6},
%  number={1},
%  pages={1--8},
%  year={2020},
%  publisher={Nature Publishing Group}
%}
%

\acknowledgements
We acknowledge support from the Deutsche Forschungsgemeinschaft (DFG, German Research Foundation) project number BR 5221/4-1, the  MSCA project pERFEcTO (Grant No. 795782), the H2020-FETOPEN-2018-2020 TEQ (grant nr. 766900), the DfE-SFI Investigator Programme (grant 15/IA/2864), COST Action CA15220, the Royal Society Wolfson Research Fellowship (RSWF\textbackslash R3\textbackslash183013), the Leverhulme Trust Research Project Grant (grant nr.~RGP-2018-266), the UK EPSRC (grant nr.~EP/T028106/1).

%\textbf{Outlooks for us for future endeavours:
%\begin{itemize}
%\item three-level systems and information driven engines
%\item what happens when also feedback control comes into the game
%\end{itemize} }

%%%%%%%%%%%%%%%%%%%%%%%%%%%%%%%%

\appendix

%%%%%%%%%%%%%%%%%%%%%%%%%%%%%%%%
%
%\appendix
\section{\label{app:incoherent_models} Classical (incoherent) CM$^2$}
%
%%%%%%%%%%%%%%%%%%%%%%%%%%%%%%%%

It is interesting to enquire what are the classical analogs of the quantum model put forth in Sec.~\ref{sec:setup}.
Or, put it differently, what are the conditions for the model to be called classical, or incoherent. 

Let us focus on a single collision event. 
We assume that, at a certain instant of time, the system is at $\rho_X = \sum_x p(x) |x\rangle\langle x |$ for some basis $|x\rangle$, while the ancilla is prepared in $\rho_Y = \sum_y p(y) |y\rangle\langle y|$, for some basis $|y\rangle$. 
The unconditional state of the system after one collision will then be
\[
\rho_X' = \mathcal{E}(\rho_X) = \sum\limits_{x,y,y'} p(x) p(y) \langle y' | U | x y\rangle\langle x y | U^\dagger | y'\rangle, 
\]
where $\langle y' | U | x y\rangle$ is still a ket in the Hilbert space of the system. 
This ket is not normalized, however, so we define 
\begin{equation}\label{state_phi}
|\Psi_{xyy'} \rangle := \frac{\langle y' | U | x y\rangle}{\sqrt{P(y'|xy)}}, 
\qquad
P(y'|xy) = || \langle y' | U | xy\rangle||^2.
\end{equation}
The state of the system may then be written as 
\[
\rho_X' = \sum\limits_{xyy'} p(x) p(y) P(y'|xy) |\Psi_{xyy'}\rangle\langle \Psi_{xyy'}|. 
\]
When written in this way, it gives the impression that $\rho_X'$ is already in diagonal form. 
But this is not the case, since in general the states $|\Psi_{xyy'}\rangle$ are not orthogonal and do not form a basis. 
Moreover, there are usually many more states than that required to span the Hilbert space of $X$ (there can be up to $d_X d_Y^2$ of them, where $d_X$, $d_Y$ are the dimensions of system and ancilla). 
As a matter of fact, in general the eigenvectors of $\rho_X'$ will have no simple relation with the states $|\Psi_{xyy'}\rangle$. 

Conversely, we say a model is \emph{unconditionally incoherent} if for any $xyy'$, the states $|\Psi_{xyy'}\rangle$ are always elements of the basis $|x\rangle$. 
In this case $\rho_X'$ will be automatically diagonal, 
\begin{equation}
\rho_X' = \sum\limits_{x'} p(x') |x'\rangle\langle x' | ,
\end{equation}
where the populations $p(x')$ can be found from 
\[
p(x') = \langle x' | \rho_X' | x' \rangle = \sum\limits_{x,y,y'} p(x) p(y) P(y'|xy) \langle x' | \Psi_{xyy'} \rangle \langle \Psi_{xyy'} | x'\rangle. 
\]
Using~\eqref{state_phi}, we can also write this as 
\begin{equation}
p(x') = \sum\limits_{x,y,y'} Q(x'y'|xy) p(x) p(y) , 
\end{equation}
where 
\begin{equation}\label{Q}
Q(x'y'|xy) = |\langle x' y' | U | x y \rangle|^2, 
\end{equation}
is the transition probability of observing a transition $(x,y) \to (x',y')$. 
A matrix of this form is said to be unistochastic, which is a particular case of doubly stochastic matrices.

An example of a unconditionally incoherent model is when both system and ancillae are qubits, interacting with the partial SWAP 
\begin{IEEEeqnarray}{rCl}
U &=& \Big(|00\rangle\langle 00 | + |11\rangle\langle 11| \Big)  \\[0.2cm]
\nonumber
&&+ \lambda \Big(|01\rangle\langle 01| + |10\rangle\langle 10|\Big)-i \sqrt{1-\lambda^2} \Big( |01 \rangle\langle 10| + |10\rangle\langle 01| \Big). 
\end{IEEEeqnarray}
In this case 
\begin{equation}
Q = \begin{pmatrix}
1 & 0 & 0 & 0 \\
0 & \lambda^2 & 1-\lambda^2 & 0 \\
0 & 1- \lambda^2 & \lambda^2 & 0 \\
0 & 0 & 0 & 1 
\end{pmatrix},
\end{equation}
with $\lambda \in [0,1]$. 

In unconditionally incoherent models, if the system is originally diagonal in the basis $|x\rangle$, it will remain so  throughout the evolution, with the populations evolving according to the classical Markov chain 
\begin{equation}
p(x_{t+1}) = \sum\limits_{x_t} \mathcal{Q}(x_{t+1}| x_t) p(x_t), 
\qquad 
\mathcal{Q}(x'|x) = \sum\limits_{y,y'} Q(x'y'|xy) p(y). 
\end{equation}

Next we can do the same for the conditional map $\mathcal{E}_z$ in Eq.~\eqref{conditional_map_def}. 
As we will see, however, unconditional incoherence does not imply conditional incoherence. 
Following the same steps as before, we can write 
\[
\mathcal{E}_z( \rho_X) = \sum\limits_{xyy'} p(x) p(y) \langle y' | M_z U |x y \rangle\langle xy | U^\dagger M_z^\dagger | y'\rangle. 
\]
We now introduce two completeness relations in the $y$ basis:
\begin{widetext}
\[
\mathcal{E}_z( \rho_X) = \sum\limits_{xyy'y''y'''} p(x) p(y) \langle y' | M_z | y''\rangle \langle y''| U |x y \rangle\langle xy | U^\dagger | y'''\rangle\langle y'''| M_z^\dagger | y'\rangle. 
\]
\end{widetext}
If the model is unconditionally incoherent, the states $\langle y''| U |x y \rangle$ will  be elements of the basis $|x\rangle$. 
But the resulting state will in general not be diagonal due to the terms $\langle y' | M_z | y''\rangle$ and $\langle y''' | M_z | y'\rangle$. 
In other words, coherence may very well be produced by the measurement itself. 
And while this cannot affect the unconditional dynamics of the system (due to no-signaling), it may very well affect the conditional one.  

We therefore define a model to be \emph{conditionally incoherent} if it is unconditionally incoherent \emph{and} if 
\[
\langle y' | M_z | y'' \rangle \propto \Delta_{y', y''}.
\] 
The simplest possibility would, of course, be to take $M_z$ as projective measurements in the basis $|y\rangle$. 
But there may also be other interesting possibilities. For instance, we can take $M_z$ to be an imprecise projective measurement, which only runs over certain elements of the basis $|y\rangle$. 
Or we could make $M_z$ be a noisy measurement, that blurs the outcomes of each $|y\rangle$. 
It is worth noting, in passing, that conditional incoherence also  immediately implies the validity of Eq.~\eqref{phi_c_uc} on the  entropy fluxes for conditionally incoherent models.

In any case, when the model is conditionally incoherent the map~\eqref{conditional_map_def} can be written as  
%\begin{equation}\label{conditional_map_elements_incoherent}
%\mathcal{E}_z(\rho_X) = \sum\limits_{x,y,y'} p(x) p(y) M(z|y') P(y'|xy) \langle y' | U | x y \rangle \langle x y | U^\dagger | y'\rangle ,
%\end{equation} 
%where 
%\begin{equation}\label{M}
%M(z|y') = \langle y' | M_z | y' \rangle, 
%\end{equation} 
%is the conditional probability of observing outcome $z$, given that the ancilla is in $|y'\rangle$. 
%\textcolor{blue}{\textbf{Ale: I think here there is something wrong, I would say it should be}\\
\begin{equation}\label{conditional_map_elements_incoherent}
\mathcal{E}_z(\rho_X) = \sum\limits_{x,y,y'} p(x) p(y) M(z|y') P(y'|xy) | \Psi_{xyy'} \rangle \langle \Psi_{xyy'} | ,
\end{equation} 
where 
\begin{equation}\label{M}
M(z|y') =| \langle y' | M_z | y' \rangle |^2={\rm tr}[M_z^\dag M_z| y' \rangle\langle y'|], 
\end{equation} 
is the conditional probability of observing outcome $z$, given that the ancilla is in $|y'\rangle$.  
This therefore represents the ``post-processing'' of the ancillary state.
The state~\eqref{conditional_map_elements_incoherent} can also be written as  
\begin{equation}\label{conditional_map_elements_incoherent_short}
\mathcal{E}_z(\rho_X) = \sum\limits_{x'} p(x',z) |x'\rangle\langle x' |, 
\end{equation}
where
\[
p(x',z) = \sum\limits_{x,y,y'} p(x) p(y) M(z|y') Q(x'y'|xy). 
\]
This is consistent with Eq.~\eqref{normalization_Pzeta}: since the result of the map is a distribution in both $x'$ and $z$, if we trace over $X$ we are left only with $p(z)$. 

At this point it is convenient to define the transition matrix
\begin{equation}\label{W}
W(x' z| x) = \sum\limits_{y,y'} M(z|y') Q(x'y'|xy) p(y). 
\end{equation}
In a classical context, this is the most important object defining a $\text{CM}^2$. 
It describes the (Markovian) transition probability, of observing the system in $x'$, as well as the outcome $z$, given that initially the system was in $x$. 
With this definition, it follows that 
\[
p(x',z) = \sum\limits_x W(x'z|x) p(x), 
\]
which, classically, is precisely what one would expect from the law of total probability. 

Finally, we adapt these ideas to multiple collisions. 
The initial state of the system is $\rho_{X_0} = \sum_{x_0} p(x_0) |x_0\rangle\langle x_0 | $. 
The conditional (unnormalized) state after the first collision is obtained by applying~\eqref{conditional_map_elements_incoherent_short}:
\[
\varrho_{X_1|\zeta_1} = \sum\limits_{x_1} p(x_1, \zeta_1) |x_1\rangle\langle x_1 | , 
\quad 
p(x_1, \zeta_1) = \sum\limits_{x_0} W(x_1 z_1| x_0)p(x_0),
\]
where, recall $\zeta_1 = z_1$. 
%\textcolor{blue}{\textbf{Ale:  missing a $p(x_0)$}
%\[
%p(x_1, \zeta_1) = \sum\limits_{x_0} W(x_1 z_0| x_0)p(x_0),
%\]
%}
Similarly, after the second collision, the conditional state will be $\varrho_{X_2|\zeta_2} = \sum\limits_{x_2} p(x_2,\zeta_2) |x_2\rangle\langle x_2 |$, where 
\[
p(x_2,\zeta_2) = \sum\limits_{x_0,x_1} W(x_2 z_2| x_1) W(x_1 z_1|x_0) p(x_0), 
\]
Proceeding in this way, we then see that after the $t$-th collision, the state of the conditional system will then be 
\begin{equation}
\varrho_{X_t|\zeta_t} = \sum\limits_{x_t} p(x_t,\zeta_t) |x_t\rangle\langle x_t |, 
\end{equation}
where
\begin{equation}
p(x_t,\zeta_t) = \sum\limits_{x_0, \ldots, x_{t-1}} W(x_t z_{t-1}| x_{t-1}) \ldots W(x_1 z_0|x_0) p(x_0), 
\end{equation}
Tracing over this state and recalling Eq.~\eqref{normalization_Pzeta}, we then  finally obtain the distribution of outcomes
\begin{equation}
P(\zeta_t) = \sum\limits_{x_0,\ldots, x_t} W(x_t z_{t}| x_{t-1}) \ldots W(x_1 z_1|x_0) p(x_0)
\end{equation}
This result is quite important, as it clearly highlights the  hidden Markov structure of the present model, discussed in Sec.~\ref{sec:setup}. 
 
Summarizing, the incoherent version of a $\text{CM}^2$ is completely defined by the transition matrix $W(x'z|x)$ in Eq.~\eqref{W}. 
This, in turn, depends on the transition matrix $Q(x'y'|xy)$ in Eq.~\eqref{Q}, which must be unistochastic, and the noise matrix $M(z'|y)$, which can be any conditional probability.  

%%%%%%%%%%%%%%%%%%%%%
\section{\label{app:proof_2ndlaw_conditional} Proof of the conditional version of the $2^\text{nd}$ law}

The proof of Eq.~\eqref{delta_sigma_c_inequality} relies on a fundamental inequality of the Holevo information~\cite{Nielsen}: 
\begin{equation}\label{holevo_ineq}
I(X' \! : \! z) \leqslant \mathcal{I}(X' \! : \! Y'). 
\end{equation}
It compares the Holevo information for a single collision outcome $z$, with the full quantum mutual information between system and ancilla, after the collision.
This means that, no matter what measurement strategy $\{M_z\}$ one utilizes, the information about the system that can be extracted from the ancilla is at most equal to the full information encoded in the global quantum state $\rho_{X'Y'}$. 
This inequality also holds for states conditioned on past outcomes. That is, 
\begin{equation}\label{holevo_ineq_2}
G_t = I_c(X_t \! : \! z_{t} | \zeta_{t-1}) \leqslant \mathcal{I}(X_t \! : \! Y_{t}' | \zeta_{t-1}), 
\end{equation}
where the conditioning is over previous records $\zeta_{t-1} = (z_1, \ldots, z_{t-1})$~(i.e., those that happened before the present collision) and $G_t$ is defined in Eq.~\eqref{gain}. 
This is true since conditional states are still quantum states (provided they are properly normalized), so that Eq.~\eqref{holevo_ineq} must still hold.
 
We now start with Eq.~\eqref{two_sigmas} and introduce the splitting~\eqref{Delta_I_splitting} to write $\Delta \Sigma_t^c = \Delta \Sigma_t^u - G_t + L_t$. 
Next we use Eq.~\eqref{sigma} for $\Delta\Sigma_t^u$ and Eq.~\eqref{gain} for $G_t$. 
We then get
\[
\Delta \Sigma_t^c = \mathcal{I}(X_t \!: \! Y_{t}') + D(Y_{t}' || Y_{t}) - I_c (X_t \! : \! z_{t} | \zeta_{t-1}) + L_t. 
\]
Using the inequality~\eqref{holevo_ineq_2} then shows that 
\begin{equation}\label{43814908821}
\Delta \Sigma_t^c \geqslant \mathcal{I}(X_t \!: \! Y_{t}') + D(Y_{t}' || Y_{t}) - \mathcal{I} (X_t \! : \! Y_{t}' | \zeta_{t-1}) + L_t. 
\end{equation}
Finally, we use Eq.~\eqref{MI_sum_entropies} for $\mathcal{I}(X_t \!: \! Y_{t}')$. 
The other mutual information $\mathcal{I} (X_t \! : \! Y_{t}' | \zeta_{t-1})$ also satisfies a similar formula 
\[
\mathcal{I} (X_t \! : \! Y_{t}' | \zeta_{t-1}) = S(X_{t} | \zeta_{t-1}) - S(X_{t-1} | \zeta_{t-1}) + S(Y_{t}' | \zeta_{t-1}) - S(Y_{t}). 
\]
Thus, the difference between the two mutual informations can be written as 
\begin{widetext}
\begin{IEEEeqnarray*}{rCl}
\mathcal{I}(X_t \!: \! Y_{t}') - \mathcal{I} (X_t \! : \! Y_{t}' | \zeta_{t-1})
&=& 
\Big[S(X_t) - S(X_t | \zeta_{t-1}) \Big]
- \Big[ S(X_{t-1}) - S(X_{t-1} | \zeta_{t-1}) \Big] 
+ \Big[S(Y_{t}') - S(Y_{t}' | \zeta_{t-1}) \Big] 
\\[0.2cm]
&=& -L_t + I(Y_{t}' \! : \! \zeta_{t-1}) ,
\end{IEEEeqnarray*}
where we recognize, in the first two square brackets, the information loss term $L_t$ defined in Eq.~\eqref{loss}.
\end{widetext}
Plugging this back in Eq.~\eqref{43814908821} we then finally find Eq.~\eqref{delta_sigma_c_inequality}. 
Being a  consequence of~\eqref{holevo_ineq_2},  we can also conclude that the first bound in~\eqref{delta_sigma_c_inequality} is saturated by processes where the measurement extracts all the information available. Even in such limiting case, we still get a non-zero $\Delta \Sigma_t^c$, so the process is still irreversible.

\bibliography{library}

\begin{thebibliography}{68}%
\makeatletter
\providecommand \@ifxundefined [1]{%
 \@ifx{#1\undefined}
}%
\providecommand \@ifnum [1]{%
 \ifnum #1\expandafter \@firstoftwo
 \else \expandafter \@secondoftwo
 \fi
}%
\providecommand \@ifx [1]{%
 \ifx #1\expandafter \@firstoftwo
 \else \expandafter \@secondoftwo
 \fi
}%
\providecommand \natexlab [1]{#1}%
\providecommand \enquote  [1]{``#1''}%
\providecommand \bibnamefont  [1]{#1}%
\providecommand \bibfnamefont [1]{#1}%
\providecommand \citenamefont [1]{#1}%
\providecommand \href@noop [0]{\@secondoftwo}%
\providecommand \href [0]{\begingroup \@sanitize@url \@href}%
\providecommand \@href[1]{\@@startlink{#1}\@@href}%
\providecommand \@@href[1]{\endgroup#1\@@endlink}%
\providecommand \@sanitize@url [0]{\catcode `\\12\catcode `\$12\catcode
  `\&12\catcode `\#12\catcode `\^12\catcode `\_12\catcode `\%12\relax}%
\providecommand \@@startlink[1]{}%
\providecommand \@@endlink[0]{}%
\providecommand \url  [0]{\begingroup\@sanitize@url \@url }%
\providecommand \@url [1]{\endgroup\@href {#1}{\urlprefix }}%
\providecommand \urlprefix  [0]{URL }%
\providecommand \Eprint [0]{\href }%
\providecommand \doibase [0]{http://dx.doi.org/}%
\providecommand \selectlanguage [0]{\@gobble}%
\providecommand \bibinfo  [0]{\@secondoftwo}%
\providecommand \bibfield  [0]{\@secondoftwo}%
\providecommand \translation [1]{[#1]}%
\providecommand \BibitemOpen [0]{}%
\providecommand \bibitemStop [0]{}%
\providecommand \bibitemNoStop [0]{.\EOS\space}%
\providecommand \EOS [0]{\spacefactor3000\relax}%
\providecommand \BibitemShut  [1]{\csname bibitem#1\endcsname}%
\let\auto@bib@innerbib\@empty
%</preamble>
\bibitem [{\citenamefont {Murch}\ \emph {et~al.}(2008)\citenamefont {Murch},
  \citenamefont {Moore}, \citenamefont {Gupta},\ and\ \citenamefont
  {Stamper-Kurn}}]{Murch2008}%
  \BibitemOpen
  \bibfield  {author} {\bibinfo {author} {\bibfnamefont {Kater~W.}\
  \bibnamefont {Murch}}, \bibinfo {author} {\bibfnamefont {Kevin~L.}\
  \bibnamefont {Moore}}, \bibinfo {author} {\bibfnamefont {Subhadeep}\
  \bibnamefont {Gupta}}, \ and\ \bibinfo {author} {\bibfnamefont {Dan~M.}\
  \bibnamefont {Stamper-Kurn}},\ }\bibfield  {title} {\enquote {\bibinfo
  {title} {{Observation of quantum-measurement backaction with an ultracold
  atomic gas}},}\ }\href {\doibase 10.1038/nphys965} {\bibfield  {journal}
  {\bibinfo  {journal} {Nat. Phys.}\ }\textbf {\bibinfo {volume} {4}},\
  \bibinfo {pages} {561--564} (\bibinfo {year} {2008})},\ \Eprint
  {http://arxiv.org/abs/arXiv:0706.1005v3} {arXiv:arXiv:0706.1005v3}
  \BibitemShut {NoStop}%
\bibitem [{\citenamefont {Purdy}\ \emph {et~al.}(2013)\citenamefont {Purdy},
  \citenamefont {Peterson},\ and\ \citenamefont {Regal}}]{Purdy2013}%
  \BibitemOpen
  \bibfield  {author} {\bibinfo {author} {\bibfnamefont {T.~P.}\ \bibnamefont
  {Purdy}}, \bibinfo {author} {\bibfnamefont {R.~W.}\ \bibnamefont {Peterson}},
  \ and\ \bibinfo {author} {\bibfnamefont {C.~A.}\ \bibnamefont {Regal}},\
  }\bibfield  {title} {\enquote {\bibinfo {title} {{Observation of radiation
  pressure shot noise on a macroscopic object}},}\ }\href
  {https://science.sciencemag.org/content/339/6121/801} {\bibfield  {journal}
  {\bibinfo  {journal} {Science}\ }\textbf {\bibinfo {volume} {339}},\ \bibinfo
  {pages} {801} (\bibinfo {year} {2013})}\BibitemShut {NoStop}%
\bibitem [{\citenamefont {Teufel}\ \emph {et~al.}(2016)\citenamefont {Teufel},
  \citenamefont {Lecocq},\ and\ \citenamefont {Simmonds}}]{Teufel2016}%
  \BibitemOpen
  \bibfield  {author} {\bibinfo {author} {\bibfnamefont {J.}~\bibnamefont
  {Teufel}}, \bibinfo {author} {\bibfnamefont {F.}~\bibnamefont {Lecocq}}, \
  and\ \bibinfo {author} {\bibfnamefont {R.}~\bibnamefont {Simmonds}},\
  }\bibfield  {title} {\enquote {\bibinfo {title} {{Overwhelming
  thermomechanical motion with microwave radiation pressure shot noise}},}\
  }\href {\doibase 10.1103/PhysRevLett.116.013602} {\bibfield  {journal}
  {\bibinfo  {journal} {Phys. Rev. Lett.}\ }\textbf {\bibinfo {volume} {116}},\
  \bibinfo {pages} {013602} (\bibinfo {year} {2016})}\BibitemShut {NoStop}%
\bibitem [{\citenamefont {Minev}\ \emph {et~al.}(2019)\citenamefont {Minev},
  \citenamefont {Mundhada}, \citenamefont {Shankar}, \citenamefont {Reinhold},
  \citenamefont {Gutierrez-Jauregui}, \citenamefont {Schoelkopf}, \citenamefont
  {Mirrahimi}, \citenamefont {Carmichael},\ and\ \citenamefont
  {Devoret}}]{Minev2018}%
  \BibitemOpen
  \bibfield  {author} {\bibinfo {author} {\bibfnamefont {Z.~K.}\ \bibnamefont
  {Minev}}, \bibinfo {author} {\bibfnamefont {S.~O.}\ \bibnamefont {Mundhada}},
  \bibinfo {author} {\bibfnamefont {S.}~\bibnamefont {Shankar}}, \bibinfo
  {author} {\bibfnamefont {P.}~\bibnamefont {Reinhold}}, \bibinfo {author}
  {\bibfnamefont {R.}~\bibnamefont {Gutierrez-Jauregui}}, \bibinfo {author}
  {\bibfnamefont {R.~J.}\ \bibnamefont {Schoelkopf}}, \bibinfo {author}
  {\bibfnamefont {M.}~\bibnamefont {Mirrahimi}}, \bibinfo {author}
  {\bibfnamefont {H.~J.}\ \bibnamefont {Carmichael}}, \ and\ \bibinfo {author}
  {\bibfnamefont {M.~H.}\ \bibnamefont {Devoret}},\ }\bibfield  {title}
  {\enquote {\bibinfo {title} {{To catch and reverse a quantum jump
  mid-flight}},}\ }\href {\doibase 10.1038/s41586-019-1287-z} {\bibfield
  {journal} {\bibinfo  {journal} {Nature}\ }\textbf {\bibinfo {volume} {570}},\
  \bibinfo {pages} {200--204} (\bibinfo {year} {2019})},\ \Eprint
  {http://arxiv.org/abs/1803.00545} {arXiv:1803.00545} \BibitemShut {NoStop}%
\bibitem [{\citenamefont {Binder}\ \emph {et~al.}(2019)\citenamefont {Binder},
  \citenamefont {Correa}, \citenamefont {Gogolin}, \citenamefont {Anders},\
  and\ \citenamefont {Adesso}}]{Binder2018a}%
  \BibitemOpen
  \bibinfo {editor} {\bibfnamefont {F.}~\bibnamefont {Binder}}, \bibinfo
  {editor} {\bibfnamefont {L.~A.}\ \bibnamefont {Correa}}, \bibinfo {editor}
  {\bibfnamefont {C.}~\bibnamefont {Gogolin}}, \bibinfo {editor} {\bibfnamefont
  {J.}~\bibnamefont {Anders}}, \ and\ \bibinfo {editor} {\bibfnamefont
  {G}~\bibnamefont {Adesso}},\ eds.,\ \href {\doibase
  10.1007/978-3-319-99046-0} {\emph {\bibinfo {title} {{Thermodynamics in the
  Quantum Regime - Fundamental Aspects and New Directions}}}}\ (\bibinfo
  {publisher} {Springer International Publishing},\ \bibinfo {address}
  {Switzerland},\ \bibinfo {year} {2019})\ p.\ \bibinfo {pages}
  {976}\BibitemShut {NoStop}%
\bibitem [{\citenamefont {Naghiloo}\ \emph {et~al.}(2018)\citenamefont
  {Naghiloo}, \citenamefont {Alonso}, \citenamefont {Romito}, \citenamefont
  {Lutz},\ and\ \citenamefont {Murch}}]{Naghiloo2018}%
  \BibitemOpen
  \bibfield  {author} {\bibinfo {author} {\bibfnamefont {M.}~\bibnamefont
  {Naghiloo}}, \bibinfo {author} {\bibfnamefont {J.~J.}\ \bibnamefont
  {Alonso}}, \bibinfo {author} {\bibfnamefont {A.}~\bibnamefont {Romito}},
  \bibinfo {author} {\bibfnamefont {E.}~\bibnamefont {Lutz}}, \ and\ \bibinfo
  {author} {\bibfnamefont {K.~W.}\ \bibnamefont {Murch}},\ }\bibfield  {title}
  {\enquote {\bibinfo {title} {{Information Gain and Loss for a Quantum
  Maxwell's Demon}},}\ }\href {\doibase 10.1103/PhysRevLett.121.030604}
  {\bibfield  {journal} {\bibinfo  {journal} {Phys. Rev. Lett.}\ }\textbf
  {\bibinfo {volume} {121}},\ \bibinfo {pages} {030604} (\bibinfo {year}
  {2018})},\ \Eprint {http://arxiv.org/abs/1802.07205} {arXiv:1802.07205}
  \BibitemShut {NoStop}%
\bibitem [{\citenamefont {Rossi}\ \emph {et~al.}(2019)\citenamefont {Rossi},
  \citenamefont {Mason}, \citenamefont {Chen},\ and\ \citenamefont
  {Schliesser}}]{Rossi2019}%
  \BibitemOpen
  \bibfield  {author} {\bibinfo {author} {\bibfnamefont {Massimiliano}\
  \bibnamefont {Rossi}}, \bibinfo {author} {\bibfnamefont {David}\ \bibnamefont
  {Mason}}, \bibinfo {author} {\bibfnamefont {Junxin}\ \bibnamefont {Chen}}, \
  and\ \bibinfo {author} {\bibfnamefont {Albert}\ \bibnamefont {Schliesser}},\
  }\bibfield  {title} {\enquote {\bibinfo {title} {{Observing and Verifying the
  Quantum Trajectory of a Mechanical Resonator}},}\ }\href {\doibase
  10.1103/PhysRevLett.123.163601} {\bibfield  {journal} {\bibinfo  {journal}
  {Phys. Rev. Lett.}\ }\textbf {\bibinfo {volume} {123}},\ \bibinfo {pages}
  {163601} (\bibinfo {year} {2019})},\ \Eprint
  {http://arxiv.org/abs/1812.00928} {arXiv:1812.00928} \BibitemShut {NoStop}%
\bibitem [{\citenamefont {Sagawa}\ and\ \citenamefont
  {Ueda}(2008)}]{Sagawa2008}%
  \BibitemOpen
  \bibfield  {author} {\bibinfo {author} {\bibfnamefont {Takahiro}\
  \bibnamefont {Sagawa}}\ and\ \bibinfo {author} {\bibfnamefont {Masahito}\
  \bibnamefont {Ueda}},\ }\bibfield  {title} {\enquote {\bibinfo {title}
  {{Second law of thermodynamics with discrete quantum feedback control}},}\
  }\href {\doibase 10.1103/PhysRevLett.100.080403} {\bibfield  {journal}
  {\bibinfo  {journal} {Phys. Rev. Lett.}\ }\textbf {\bibinfo {volume} {100}},\
  \bibinfo {pages} {080403} (\bibinfo {year} {2008})},\ \Eprint
  {http://arxiv.org/abs/0710.0956} {arXiv:0710.0956} \BibitemShut {NoStop}%
\bibitem [{\citenamefont {Ito}\ and\ \citenamefont {Sagawa}(2013)}]{Ito2013}%
  \BibitemOpen
  \bibfield  {author} {\bibinfo {author} {\bibfnamefont {Sosuke}\ \bibnamefont
  {Ito}}\ and\ \bibinfo {author} {\bibfnamefont {Takahiro}\ \bibnamefont
  {Sagawa}},\ }\bibfield  {title} {\enquote {\bibinfo {title} {{Information
  Thermodynamics on Causal Networks}},}\ }\href {\doibase
  10.1103/PhysRevLett.111.180603} {\bibfield  {journal} {\bibinfo  {journal}
  {Phys. Rev. Lett.}\ }\textbf {\bibinfo {volume} {111}},\ \bibinfo {pages}
  {180603} (\bibinfo {year} {2013})},\ \Eprint {http://arxiv.org/abs/1306.2756}
  {arXiv:1306.2756} \BibitemShut {NoStop}%
\bibitem [{\citenamefont {Sagawa}\ and\ \citenamefont
  {Ueda}(2012)}]{Sagawa2012}%
  \BibitemOpen
  \bibfield  {author} {\bibinfo {author} {\bibfnamefont {Takahiro}\
  \bibnamefont {Sagawa}}\ and\ \bibinfo {author} {\bibfnamefont {Masahito}\
  \bibnamefont {Ueda}},\ }\bibfield  {title} {\enquote {\bibinfo {title}
  {{Fluctuation Theorem with Information Exchange: Role of Correlations in
  Stochastic Thermodynamics}},}\ }\href {\doibase
  10.1103/PhysRevLett.109.180602} {\bibfield  {journal} {\bibinfo  {journal}
  {Physical Review Letters}\ }\textbf {\bibinfo {volume} {109}},\ \bibinfo
  {pages} {180602} (\bibinfo {year} {2012})}\BibitemShut {NoStop}%
\bibitem [{\citenamefont {Sagawa}\ and\ \citenamefont
  {Ueda}(2013)}]{Sagawa2013}%
  \BibitemOpen
  \bibfield  {author} {\bibinfo {author} {\bibfnamefont {Takahiro}\
  \bibnamefont {Sagawa}}\ and\ \bibinfo {author} {\bibfnamefont {Masahito}\
  \bibnamefont {Ueda}},\ }\bibfield  {title} {\enquote {\bibinfo {title} {{Role
  of mutual information in entropy production under information exchanges}},}\
  }\href {\doibase 10.1088/1367-2630/15/12/125012} {\bibfield  {journal}
  {\bibinfo  {journal} {New J. Phys.}\ }\textbf {\bibinfo {volume} {15}},\
  \bibinfo {pages} {125012} (\bibinfo {year} {2013})}\BibitemShut {NoStop}%
\bibitem [{\citenamefont {Funo}\ \emph {et~al.}(2013)\citenamefont {Funo},
  \citenamefont {Watanabe},\ and\ \citenamefont {Ueda}}]{Funo2013}%
  \BibitemOpen
  \bibfield  {author} {\bibinfo {author} {\bibfnamefont {Ken}\ \bibnamefont
  {Funo}}, \bibinfo {author} {\bibfnamefont {Yu}~\bibnamefont {Watanabe}}, \
  and\ \bibinfo {author} {\bibfnamefont {Masahito}\ \bibnamefont {Ueda}},\
  }\bibfield  {title} {\enquote {\bibinfo {title} {{Integral quantum
  fluctuation theorems under measurement and feedback control}},}\ }\href
  {\doibase 10.1103/PhysRevE.88.052121} {\bibfield  {journal} {\bibinfo
  {journal} {Phys. Rev. E}\ }\textbf {\bibinfo {volume} {88}},\ \bibinfo
  {pages} {052121} (\bibinfo {year} {2013})},\ \Eprint
  {http://arxiv.org/abs/1307.2362} {arXiv:1307.2362} \BibitemShut {NoStop}%
\bibitem [{\citenamefont {Elouard}\ \emph {et~al.}(2017)\citenamefont
  {Elouard}, \citenamefont {Herrera-Mart{\'{i}}}, \citenamefont {Clusel},\ and\
  \citenamefont {Auff{\`{e}}ves}}]{Elouard2017a}%
  \BibitemOpen
  \bibfield  {author} {\bibinfo {author} {\bibfnamefont {Cyril}\ \bibnamefont
  {Elouard}}, \bibinfo {author} {\bibfnamefont {David~A.}\ \bibnamefont
  {Herrera-Mart{\'{i}}}}, \bibinfo {author} {\bibfnamefont {Maxime}\
  \bibnamefont {Clusel}}, \ and\ \bibinfo {author} {\bibfnamefont {Alexia}\
  \bibnamefont {Auff{\`{e}}ves}},\ }\bibfield  {title} {\enquote {\bibinfo
  {title} {{The role of quantum measurement in stochastic thermodynamics}},}\
  }\href {\doibase 10.1038/s41534-017-0008-4} {\bibfield  {journal} {\bibinfo
  {journal} {npj Quant. Inf.}\ }\textbf {\bibinfo {volume} {3}},\ \bibinfo
  {pages} {9} (\bibinfo {year} {2017})},\ \Eprint
  {http://arxiv.org/abs/1607.02404} {arXiv:1607.02404} \BibitemShut {NoStop}%
\bibitem [{\citenamefont {Buffoni}\ \emph {et~al.}(2018)\citenamefont
  {Buffoni}, \citenamefont {Solfanelli}, \citenamefont {Verrucchi},
  \citenamefont {Cuccoli},\ and\ \citenamefont {Campisi}}]{Buffoni2018}%
  \BibitemOpen
  \bibfield  {author} {\bibinfo {author} {\bibfnamefont {Lorenzo}\ \bibnamefont
  {Buffoni}}, \bibinfo {author} {\bibfnamefont {Andrea}\ \bibnamefont
  {Solfanelli}}, \bibinfo {author} {\bibfnamefont {Paola}\ \bibnamefont
  {Verrucchi}}, \bibinfo {author} {\bibfnamefont {Alessandro}\ \bibnamefont
  {Cuccoli}}, \ and\ \bibinfo {author} {\bibfnamefont {Michele}\ \bibnamefont
  {Campisi}},\ }\bibfield  {title} {\enquote {\bibinfo {title} {{Quantum
  Measurement Cooling}},}\ }\href {\doibase 10.1103/PhysRevLett.122.070603}
  {\bibfield  {journal} {\bibinfo  {journal} {Physical Review Letters}\
  }\textbf {\bibinfo {volume} {122}},\ \bibinfo {pages} {070603} (\bibinfo
  {year} {2018})},\ \Eprint {http://arxiv.org/abs/1806.07814}
  {arXiv:1806.07814} \BibitemShut {NoStop}%
\bibitem [{\citenamefont {Mohammady}\ and\ \citenamefont
  {Romito}(2019)}]{Mohammady2019a}%
  \BibitemOpen
  \bibfield  {author} {\bibinfo {author} {\bibfnamefont {M.~Hamed}\
  \bibnamefont {Mohammady}}\ and\ \bibinfo {author} {\bibfnamefont
  {Alessandro}\ \bibnamefont {Romito}},\ }\bibfield  {title} {\enquote
  {\bibinfo {title} {{Conditional work statistics of quantum measurements}},}\
  }\href {\doibase 10.22331/q-2019-08-19-175} {\bibfield  {journal} {\bibinfo
  {journal} {Quantum}\ }\textbf {\bibinfo {volume} {3}},\ \bibinfo {pages}
  {175} (\bibinfo {year} {2019})},\ \Eprint {http://arxiv.org/abs/1809.09010}
  {arXiv:1809.09010} \BibitemShut {NoStop}%
\bibitem [{\citenamefont {Beyer}\ \emph {et~al.}(2020)\citenamefont {Beyer},
  \citenamefont {Luoma},\ and\ \citenamefont {Strunz}}]{Beyer2020}%
  \BibitemOpen
  \bibfield  {author} {\bibinfo {author} {\bibfnamefont {Konstantin}\
  \bibnamefont {Beyer}}, \bibinfo {author} {\bibfnamefont {Kimmo}\ \bibnamefont
  {Luoma}}, \ and\ \bibinfo {author} {\bibfnamefont {Walter~T.}\ \bibnamefont
  {Strunz}},\ }\bibfield  {title} {\enquote {\bibinfo {title} {{Work as an
  external quantum observable and an operational quantum work fluctuation
  theorem}},}\ }\href {\doibase 10.1103/physrevresearch.2.033508} {\bibfield
  {journal} {\bibinfo  {journal} {Phys. Rev. Research}\ }\textbf {\bibinfo
  {volume} {2}},\ \bibinfo {pages} {33508} (\bibinfo {year}
  {2020})}\BibitemShut {NoStop}%
\bibitem [{\citenamefont {Sone}\ and\ \citenamefont
  {Deffner}(2020)}]{Sone2020}%
  \BibitemOpen
  \bibfield  {author} {\bibinfo {author} {\bibfnamefont {Akira}\ \bibnamefont
  {Sone}}\ and\ \bibinfo {author} {\bibfnamefont {Sebastian}\ \bibnamefont
  {Deffner}},\ }\bibfield  {title} {\enquote {\bibinfo {title} {{Jarzynski
  equality for conditional stochastic work}},}\ }\href
  {http://arxiv.org/abs/2010.05835} {\  (\bibinfo {year} {2020})},\ \Eprint
  {http://arxiv.org/abs/2010.05835} {arXiv:2010.05835} \BibitemShut {NoStop}%
\bibitem [{\citenamefont {Strasberg}\ and\ \citenamefont
  {Winter}(2019)}]{Strasberg2019a}%
  \BibitemOpen
  \bibfield  {author} {\bibinfo {author} {\bibfnamefont {Philipp}\ \bibnamefont
  {Strasberg}}\ and\ \bibinfo {author} {\bibfnamefont {Andreas}\ \bibnamefont
  {Winter}},\ }\bibfield  {title} {\enquote {\bibinfo {title} {{Stochastic
  thermodynamics with arbitrary interventions}},}\ }\href {\doibase
  10.1103/PhysRevE.100.022135} {\bibfield  {journal} {\bibinfo  {journal}
  {Phys. Rev. E}\ }\textbf {\bibinfo {volume} {100}},\ \bibinfo {pages}
  {022135} (\bibinfo {year} {2019})},\ \Eprint
  {http://arxiv.org/abs/1905.07990} {arXiv:1905.07990} \BibitemShut {NoStop}%
\bibitem [{\citenamefont {Strasberg}(2020)}]{Strasberg2020}%
  \BibitemOpen
  \bibfield  {author} {\bibinfo {author} {\bibfnamefont {Philipp}\ \bibnamefont
  {Strasberg}},\ }\bibfield  {title} {\enquote {\bibinfo {title}
  {{Thermodynamics of Quantum Causal Models: An Inclusive, Hamiltonian
  Approach}},}\ }\href {\doibase 10.22331/q-2020-03-02-240} {\bibfield
  {journal} {\bibinfo  {journal} {Quantum}\ }\textbf {\bibinfo {volume} {4}},\
  \bibinfo {pages} {240} (\bibinfo {year} {2020})},\ \Eprint
  {http://arxiv.org/abs/1911.01730} {arXiv:1911.01730} \BibitemShut {NoStop}%
\bibitem [{\citenamefont {Belenchia}\ \emph {et~al.}(2020)\citenamefont
  {Belenchia}, \citenamefont {Mancino}, \citenamefont {Landi},\ and\
  \citenamefont {Paternostro}}]{Belenchia2019}%
  \BibitemOpen
  \bibfield  {author} {\bibinfo {author} {\bibfnamefont {Alessio}\ \bibnamefont
  {Belenchia}}, \bibinfo {author} {\bibfnamefont {Luca}\ \bibnamefont
  {Mancino}}, \bibinfo {author} {\bibfnamefont {Gabriel~T.}\ \bibnamefont
  {Landi}}, \ and\ \bibinfo {author} {\bibfnamefont {Mauro}\ \bibnamefont
  {Paternostro}},\ }\bibfield  {title} {\enquote {\bibinfo {title} {{Entropy
  Production in Continuously Measured Quantum Systems}},}\ }\href {\doibase
  10.1038/s41534-020-00334-6} {\bibfield  {journal} {\bibinfo  {journal} {npj
  Quant. Inf.}\ }\textbf {\bibinfo {volume} {6}},\ \bibinfo {pages} {97}
  (\bibinfo {year} {2020})},\ \Eprint {http://arxiv.org/abs/1908.09382}
  {arXiv:1908.09382} \BibitemShut {NoStop}%
\bibitem [{\citenamefont {Toyabe}\ \emph {et~al.}(2010)\citenamefont {Toyabe},
  \citenamefont {Sagawa}, \citenamefont {Ueda}, \citenamefont {Muneyuki},\ and\
  \citenamefont {Sano}}]{Toyabe2010}%
  \BibitemOpen
  \bibfield  {author} {\bibinfo {author} {\bibfnamefont {Shoichi}\ \bibnamefont
  {Toyabe}}, \bibinfo {author} {\bibfnamefont {Takahiro}\ \bibnamefont
  {Sagawa}}, \bibinfo {author} {\bibfnamefont {Masahito}\ \bibnamefont {Ueda}},
  \bibinfo {author} {\bibfnamefont {Eiro}\ \bibnamefont {Muneyuki}}, \ and\
  \bibinfo {author} {\bibfnamefont {Masaki}\ \bibnamefont {Sano}},\ }\bibfield
  {title} {\enquote {\bibinfo {title} {{Experimental demonstration of
  information-to-energy conversion and validation of the generalized Jarzynski
  equality}},}\ }\href {\doibase 10.1038/nphys1821} {\bibfield  {journal}
  {\bibinfo  {journal} {Nat. Phys.}\ }\textbf {\bibinfo {volume} {6}},\
  \bibinfo {pages} {988} (\bibinfo {year} {2010})},\ \Eprint
  {http://arxiv.org/abs/1009.5287} {arXiv:1009.5287} \BibitemShut {NoStop}%
\bibitem [{\citenamefont {Koski}\ \emph {et~al.}(2014)\citenamefont {Koski},
  \citenamefont {Maisi}, \citenamefont {Pekola},\ and\ \citenamefont
  {Averin}}]{Koski2014}%
  \BibitemOpen
  \bibfield  {author} {\bibinfo {author} {\bibfnamefont {J.~V.}\ \bibnamefont
  {Koski}}, \bibinfo {author} {\bibfnamefont {V.~F.}\ \bibnamefont {Maisi}},
  \bibinfo {author} {\bibfnamefont {J.~P.}\ \bibnamefont {Pekola}}, \ and\
  \bibinfo {author} {\bibfnamefont {D.~V.}\ \bibnamefont {Averin}},\ }\bibfield
   {title} {\enquote {\bibinfo {title} {{Experimental realization of a Szilard
  engine with a single electron}},}\ }\href {\doibase 10.1073/pnas.1406966111}
  {\bibfield  {journal} {\bibinfo  {journal} {Proc. Natl. Acad. Sci. U.S.A.}\
  }\textbf {\bibinfo {volume} {111}},\ \bibinfo {pages} {13786} (\bibinfo
  {year} {2014})},\ \Eprint {http://arxiv.org/abs/1402.5907} {arXiv:1402.5907}
  \BibitemShut {NoStop}%
\bibitem [{\citenamefont {Cottet}\ \emph {et~al.}(2017)\citenamefont {Cottet},
  \citenamefont {Jezouin}, \citenamefont {Bretheau}, \citenamefont
  {Campagne-Ibarcq}, \citenamefont {Ficheux}, \citenamefont {Anders},
  \citenamefont {Auff{\`{e}}ves}, \citenamefont {Azouit}, \citenamefont
  {Rouchon},\ and\ \citenamefont {Huard}}]{Cottet2017}%
  \BibitemOpen
  \bibfield  {author} {\bibinfo {author} {\bibfnamefont {N}~\bibnamefont
  {Cottet}}, \bibinfo {author} {\bibfnamefont {S}~\bibnamefont {Jezouin}},
  \bibinfo {author} {\bibfnamefont {L}~\bibnamefont {Bretheau}}, \bibinfo
  {author} {\bibfnamefont {P.}~\bibnamefont {Campagne-Ibarcq}}, \bibinfo
  {author} {\bibfnamefont {Q}~\bibnamefont {Ficheux}}, \bibinfo {author}
  {\bibfnamefont {Janet}\ \bibnamefont {Anders}}, \bibinfo {author}
  {\bibfnamefont {Alexia}\ \bibnamefont {Auff{\`{e}}ves}}, \bibinfo {author}
  {\bibfnamefont {R.}~\bibnamefont {Azouit}}, \bibinfo {author} {\bibfnamefont
  {P.}~\bibnamefont {Rouchon}}, \ and\ \bibinfo {author} {\bibfnamefont
  {B.}~\bibnamefont {Huard}},\ }\bibfield  {title} {\enquote {\bibinfo {title}
  {{Observing a quantum Maxwell demon at work}},}\ }\href {\doibase
  10.1073/pnas.1704827114} {\bibfield  {journal} {\bibinfo  {journal} {Proc.
  Natl. Acad. Sci. U.S.A}\ }\textbf {\bibinfo {volume} {114}},\ \bibinfo
  {pages} {7561--7564} (\bibinfo {year} {2017})},\ \Eprint
  {http://arxiv.org/abs/1702.05161} {arXiv:1702.05161} \BibitemShut {NoStop}%
\bibitem [{\citenamefont {Debiossac}\ \emph {et~al.}(2020)\citenamefont
  {Debiossac}, \citenamefont {Grass}, \citenamefont {Alonso}, \citenamefont
  {Lutz},\ and\ \citenamefont {Kiesel}}]{Debiossac2019}%
  \BibitemOpen
  \bibfield  {author} {\bibinfo {author} {\bibfnamefont {Maxime}\ \bibnamefont
  {Debiossac}}, \bibinfo {author} {\bibfnamefont {David}\ \bibnamefont
  {Grass}}, \bibinfo {author} {\bibfnamefont {Jose~Joaquin}\ \bibnamefont
  {Alonso}}, \bibinfo {author} {\bibfnamefont {Eric}\ \bibnamefont {Lutz}}, \
  and\ \bibinfo {author} {\bibfnamefont {Nikolai}\ \bibnamefont {Kiesel}},\
  }\bibfield  {title} {\enquote {\bibinfo {title} {{Thermodynamics of
  continuous non-Markovian feedback control}},}\ }\href {\doibase
  10.1038/s41467-020-15148-5} {\bibfield  {journal} {\bibinfo  {journal} {Nat.
  Commun.}\ }\textbf {\bibinfo {volume} {11}},\ \bibinfo {pages} {1360}
  (\bibinfo {year} {2020})},\ \Eprint {http://arxiv.org/abs/1904.04889}
  {arXiv:1904.04889} \BibitemShut {NoStop}%
\bibitem [{\citenamefont {Wiseman}\ and\ \citenamefont
  {Milburn}(2009)}]{Wiseman2009}%
  \BibitemOpen
  \bibfield  {author} {\bibinfo {author} {\bibfnamefont {H.~M.}\ \bibnamefont
  {Wiseman}}\ and\ \bibinfo {author} {\bibfnamefont {G.~J.}\ \bibnamefont
  {Milburn}},\ }\href@noop {} {\emph {\bibinfo {title} {{Quantum measurement
  and control}}}}\ (\bibinfo  {publisher} {Cambridge University Press},\
  \bibinfo {address} {New York},\ \bibinfo {year} {2009})\BibitemShut {NoStop}%
\bibitem [{\citenamefont {Jacobs}(2014)}]{Jacobs2014}%
  \BibitemOpen
  \bibfield  {author} {\bibinfo {author} {\bibfnamefont {Kurt}\ \bibnamefont
  {Jacobs}},\ }\href@noop {} {\emph {\bibinfo {title} {{Quantum measurement
  theory and its applications}}}}\ (\bibinfo  {publisher} {Cambridge University
  Press},\ \bibinfo {address} {Cambridge},\ \bibinfo {year} {2014})\BibitemShut
  {NoStop}%
\bibitem [{\citenamefont {Rossi}\ \emph {et~al.}(2020)\citenamefont {Rossi},
  \citenamefont {Mancino}, \citenamefont {Landi}, \citenamefont {Paternostro},
  \citenamefont {Schliesser},\ and\ \citenamefont {Belenchia}}]{Rossi2020}%
  \BibitemOpen
  \bibfield  {author} {\bibinfo {author} {\bibfnamefont {Massimiliano}\
  \bibnamefont {Rossi}}, \bibinfo {author} {\bibfnamefont {Luca}\ \bibnamefont
  {Mancino}}, \bibinfo {author} {\bibfnamefont {Gabriel~T.}\ \bibnamefont
  {Landi}}, \bibinfo {author} {\bibfnamefont {Mauro}\ \bibnamefont
  {Paternostro}}, \bibinfo {author} {\bibfnamefont {Albert}\ \bibnamefont
  {Schliesser}}, \ and\ \bibinfo {author} {\bibfnamefont {Alessio}\
  \bibnamefont {Belenchia}},\ }\bibfield  {title} {\enquote {\bibinfo {title}
  {{Experimental assessment of entropy production in a continuously measured
  mechanical resonator}},}\ }\href {\doibase 10.1103/PhysRevLett.125.080601}
  {\bibfield  {journal} {\bibinfo  {journal} {Phys. Rev. Lett.}\ }\textbf
  {\bibinfo {volume} {125}},\ \bibinfo {pages} {080601} (\bibinfo {year}
  {2020})},\ \Eprint {http://arxiv.org/abs/2005.03429} {arXiv:2005.03429}
  \BibitemShut {NoStop}%
\bibitem [{\citenamefont {Landi}\ and\ \citenamefont
  {Paternostro}()}]{Landi2020a}%
  \BibitemOpen
  \bibfield  {author} {\bibinfo {author} {\bibfnamefont {Gabriel~T.}\
  \bibnamefont {Landi}}\ and\ \bibinfo {author} {\bibfnamefont {Mauro}\
  \bibnamefont {Paternostro}},\ }\href {http://arxiv.org/abs/2009.07668}
  {\enquote {\bibinfo {title} {{Irreversible entropy production, from quantum
  to classical}},}\ }\Eprint {http://arxiv.org/abs/2009.07668}
  {arXiv:2009.07668} \BibitemShut {NoStop}%
\bibitem [{\citenamefont {Levy}\ and\ \citenamefont
  {Kosloff}(2014)}]{Levy2014}%
  \BibitemOpen
  \bibfield  {author} {\bibinfo {author} {\bibfnamefont {Amikam}\ \bibnamefont
  {Levy}}\ and\ \bibinfo {author} {\bibfnamefont {Ronnie}\ \bibnamefont
  {Kosloff}},\ }\bibfield  {title} {\enquote {\bibinfo {title} {{The local
  approach to quantum transport may violate the second law of
  thermodynamics}},}\ }\href {\doibase 10.1209/0295-5075/107/20004} {\bibfield
  {journal} {\bibinfo  {journal} {{EPL} (Europhysics Letters)}\ }\textbf
  {\bibinfo {volume} {107}},\ \bibinfo {pages} {20004} (\bibinfo {year}
  {2014})},\ \Eprint {http://arxiv.org/abs/1402.3825} {arXiv:1402.3825}
  \BibitemShut {NoStop}%
\bibitem [{\citenamefont {{De Chiara}}\ \emph {et~al.}(2018)\citenamefont {{De
  Chiara}}, \citenamefont {Landi}, \citenamefont {Hewgill}, \citenamefont
  {Reid}, \citenamefont {Ferraro}, \citenamefont {Roncaglia},\ and\
  \citenamefont {Antezza}}]{DeChiara2018}%
  \BibitemOpen
  \bibfield  {author} {\bibinfo {author} {\bibfnamefont {G.}~\bibnamefont {{De
  Chiara}}}, \bibinfo {author} {\bibfnamefont {G.}~\bibnamefont {Landi}},
  \bibinfo {author} {\bibfnamefont {A.}~\bibnamefont {Hewgill}}, \bibinfo
  {author} {\bibfnamefont {B.}~\bibnamefont {Reid}}, \bibinfo {author}
  {\bibfnamefont {A.}~\bibnamefont {Ferraro}}, \bibinfo {author} {\bibfnamefont
  {A.~J.}\ \bibnamefont {Roncaglia}}, \ and\ \bibinfo {author} {\bibfnamefont
  {M.}~\bibnamefont {Antezza}},\ }\bibfield  {title} {\enquote {\bibinfo
  {title} {{Reconciliation of quantum local master equations with
  thermodynamics}},}\ }\href {\doibase
  https://doi.org/10.1088/1367-2630/aaecee} {\bibfield  {journal} {\bibinfo
  {journal} {New J. Phys.}\ }\textbf {\bibinfo {volume} {20}},\ \bibinfo
  {pages} {113024} (\bibinfo {year} {2018})},\ \Eprint
  {http://arxiv.org/abs/1808.10450} {arXiv:1808.10450} \BibitemShut {NoStop}%
\bibitem [{\citenamefont {Rau}(1963)}]{Rau1963}%
  \BibitemOpen
  \bibfield  {author} {\bibinfo {author} {\bibfnamefont {Jayaseetha}\
  \bibnamefont {Rau}},\ }\bibfield  {title} {\enquote {\bibinfo {title}
  {{Relaxation phenomena in spin and harmonic oscillator systems}},}\ }\href
  {\doibase 10.1103/PhysRev.129.1880} {\bibfield  {journal} {\bibinfo
  {journal} {Phys. Rev.}\ }\textbf {\bibinfo {volume} {129}},\ \bibinfo {pages}
  {1880--1888} (\bibinfo {year} {1963})}\BibitemShut {NoStop}%
\bibitem [{\citenamefont {Scarani}\ \emph {et~al.}(2002)\citenamefont
  {Scarani}, \citenamefont {Ziman}, \citenamefont {{\v{S}}telmachovi{\v{c}}},
  \citenamefont {Gisin}, \citenamefont {Bu{\v{z}}ek},\ and\ \citenamefont
  {Bu{\v{z}}ek}}]{Scarani2002}%
  \BibitemOpen
  \bibfield  {author} {\bibinfo {author} {\bibfnamefont {Valerio}\ \bibnamefont
  {Scarani}}, \bibinfo {author} {\bibfnamefont {M{\'{a}}rio}\ \bibnamefont
  {Ziman}}, \bibinfo {author} {\bibfnamefont {Peter}\ \bibnamefont
  {{\v{S}}telmachovi{\v{c}}}}, \bibinfo {author} {\bibfnamefont {Nicolas}\
  \bibnamefont {Gisin}}, \bibinfo {author} {\bibfnamefont {Vladim{\'{i}}r}\
  \bibnamefont {Bu{\v{z}}ek}}, \ and\ \bibinfo {author} {\bibfnamefont
  {Vladim{\'{i}}r}\ \bibnamefont {Bu{\v{z}}ek}},\ }\bibfield  {title} {\enquote
  {\bibinfo {title} {{Thermalizing quantum machines: Dissipation and
  entanglement}},}\ }\href {\doibase 10.1103/PhysRevLett.88.097905} {\bibfield
  {journal} {\bibinfo  {journal} {Phys. Rev. Lett.}\ }\textbf {\bibinfo
  {volume} {88}},\ \bibinfo {pages} {097905} (\bibinfo {year} {2002})},\
  \Eprint {http://arxiv.org/abs/0110088} {arXiv:0110088 [quant-ph]}
  \BibitemShut {NoStop}%
\bibitem [{\citenamefont {Ziman}\ \emph {et~al.}(2002)\citenamefont {Ziman},
  \citenamefont {{\v{S}}telmachovi{\v{c}}}, \citenamefont {Buz{\v{z}}ek},
  \citenamefont {Hillery}, \citenamefont {Scarani},\ and\ \citenamefont
  {Gisin}}]{Ziman2002}%
  \BibitemOpen
  \bibfield  {author} {\bibinfo {author} {\bibfnamefont {M.}~\bibnamefont
  {Ziman}}, \bibinfo {author} {\bibfnamefont {P.}~\bibnamefont
  {{\v{S}}telmachovi{\v{c}}}}, \bibinfo {author} {\bibfnamefont
  {V.}~\bibnamefont {Buz{\v{z}}ek}}, \bibinfo {author} {\bibfnamefont
  {M.}~\bibnamefont {Hillery}}, \bibinfo {author} {\bibfnamefont
  {V.}~\bibnamefont {Scarani}}, \ and\ \bibinfo {author} {\bibfnamefont
  {N.}~\bibnamefont {Gisin}},\ }\bibfield  {title} {\enquote {\bibinfo {title}
  {{Diluting quantum information: An analysis of information transfer in
  system-reservoir interactions}},}\ }\href {\doibase
  10.1103/PhysRevA.65.042105} {\bibfield  {journal} {\bibinfo  {journal} {Phys.
  Rev. A}\ }\textbf {\bibinfo {volume} {65}},\ \bibinfo {pages} {042105}
  (\bibinfo {year} {2002})}\BibitemShut {NoStop}%
\bibitem [{\citenamefont {Englert}\ and\ \citenamefont
  {Morigi}(2002)}]{Englert2002}%
  \BibitemOpen
  \bibfield  {author} {\bibinfo {author} {\bibfnamefont {Berthold-Georg}\
  \bibnamefont {Englert}}\ and\ \bibinfo {author} {\bibfnamefont {Giovanna}\
  \bibnamefont {Morigi}},\ }\bibfield  {title} {\enquote {\bibinfo {title}
  {{Five Lectures On Dissipative Master Equations}},}\ }in\ \href@noop {}
  {\emph {\bibinfo {booktitle} {Coherent Evolution in Noisy Environments -
  Lecture Notes in Physics}}},\ \bibinfo {editor} {edited by\ \bibinfo {editor}
  {\bibfnamefont {A.}~\bibnamefont {Buchleitner}}\ and\ \bibinfo {editor}
  {\bibfnamefont {K.}~\bibnamefont {Hornberger}}}\ (\bibinfo  {publisher}
  {Springer},\ \bibinfo {address} {Berlin, Heidelberg},\ \bibinfo {year}
  {2002})\ p.\ \bibinfo {pages} {611},\ \Eprint {http://arxiv.org/abs/0206116}
  {arXiv:0206116 [quant-ph]} \BibitemShut {NoStop}%
\bibitem [{\citenamefont {Attal}\ and\ \citenamefont
  {Pautrat}(2006)}]{Attal2006a}%
  \BibitemOpen
  \bibfield  {author} {\bibinfo {author} {\bibfnamefont {St{\'{e}}phane}\
  \bibnamefont {Attal}}\ and\ \bibinfo {author} {\bibfnamefont {Yan}\
  \bibnamefont {Pautrat}},\ }\bibfield  {title} {\enquote {\bibinfo {title}
  {{From Repeated to Continuous Quantum Interactions}},}\ }\href {\doibase
  10.1007/s00023-005-0242-8} {\bibfield  {journal} {\bibinfo  {journal}
  {Annales Henri Poincar{\'{e}}}\ }\textbf {\bibinfo {volume} {7}},\ \bibinfo
  {pages} {59--104} (\bibinfo {year} {2006})},\ \Eprint
  {http://arxiv.org/abs/0311002} {arXiv:0311002 [math-ph]} \BibitemShut
  {NoStop}%
\bibitem [{\citenamefont {Karevski}\ and\ \citenamefont
  {Platini}(2009)}]{Karevski2009}%
  \BibitemOpen
  \bibfield  {author} {\bibinfo {author} {\bibfnamefont {D}~\bibnamefont
  {Karevski}}\ and\ \bibinfo {author} {\bibfnamefont {T}~\bibnamefont
  {Platini}},\ }\bibfield  {title} {\enquote {\bibinfo {title} {{Quantum
  Nonequilibrium Steady States Induced by Repeated Interactions}},}\ }\href
  {\doibase 10.1103/PhysRevLett.102.207207} {\bibfield  {journal} {\bibinfo
  {journal} {Phys. Rev. Lett.}\ }\textbf {\bibinfo {volume} {102}},\ \bibinfo
  {pages} {207207} (\bibinfo {year} {2009})},\ \Eprint
  {http://arxiv.org/abs/0904.3527} {arXiv:0904.3527} \BibitemShut {NoStop}%
\bibitem [{\citenamefont {Pellegrini}\ and\ \citenamefont
  {Petruccione}(2009)}]{Pellegrini2009}%
  \BibitemOpen
  \bibfield  {author} {\bibinfo {author} {\bibfnamefont {C}~\bibnamefont
  {Pellegrini}}\ and\ \bibinfo {author} {\bibfnamefont {F}~\bibnamefont
  {Petruccione}},\ }\bibfield  {title} {\enquote {\bibinfo {title}
  {{Non-Markovian quantum repeated interactions and measurements}},}\ }\href
  {\doibase 10.1088/1751-8113/42/42/425304} {\bibfield  {journal} {\bibinfo
  {journal} {J. Phys. A}\ }\textbf {\bibinfo {volume} {42}},\ \bibinfo {pages}
  {425304} (\bibinfo {year} {2009})},\ \Eprint {http://arxiv.org/abs/0903.3859}
  {arXiv:0903.3859} \BibitemShut {NoStop}%
\bibitem [{\citenamefont {Giovannetti}\ and\ \citenamefont
  {Palma}(2012)}]{Giovannetti2012}%
  \BibitemOpen
  \bibfield  {author} {\bibinfo {author} {\bibfnamefont {V.}~\bibnamefont
  {Giovannetti}}\ and\ \bibinfo {author} {\bibfnamefont {G.~M.}\ \bibnamefont
  {Palma}},\ }\bibfield  {title} {\enquote {\bibinfo {title} {{Master equations
  for correlated quantum channels}},}\ }\href {\doibase
  10.1103/PhysRevLett.108.040401} {\bibfield  {journal} {\bibinfo  {journal}
  {Phys. Rev. Lett.}\ }\textbf {\bibinfo {volume} {108}},\ \bibinfo {pages}
  {040401} (\bibinfo {year} {2012})}\BibitemShut {NoStop}%
\bibitem [{\citenamefont {Ryb{\'{a}}r}\ \emph {et~al.}(2012)\citenamefont
  {Ryb{\'{a}}r}, \citenamefont {Filippov}, \citenamefont {Ziman},\ and\
  \citenamefont {Bu{\v{z}}ek}}]{Rybar2012a}%
  \BibitemOpen
  \bibfield  {author} {\bibinfo {author} {\bibfnamefont {Tom{\'{a}}{\v{s}}}\
  \bibnamefont {Ryb{\'{a}}r}}, \bibinfo {author} {\bibfnamefont {Sergey~N.}\
  \bibnamefont {Filippov}}, \bibinfo {author} {\bibfnamefont {M{\'{a}}rio}\
  \bibnamefont {Ziman}}, \ and\ \bibinfo {author} {\bibfnamefont
  {Vladim{\'{i}}r}\ \bibnamefont {Bu{\v{z}}ek}},\ }\bibfield  {title} {\enquote
  {\bibinfo {title} {{Simulation of indivisible qubit channels in collision
  models}},}\ }\href {\doibase 10.1088/0953-4075/45/15/154006} {\bibfield
  {journal} {\bibinfo  {journal} {J. Phys. B}\ }\textbf {\bibinfo {volume}
  {45}},\ \bibinfo {pages} {154006} (\bibinfo {year} {2012})},\ \Eprint
  {http://arxiv.org/abs/1202.6315} {arXiv:1202.6315} \BibitemShut {NoStop}%
\bibitem [{\citenamefont {Strasberg}\ \emph {et~al.}(2017)\citenamefont
  {Strasberg}, \citenamefont {Schaller}, \citenamefont {Brandes},\ and\
  \citenamefont {Esposito}}]{Strasberg2016}%
  \BibitemOpen
  \bibfield  {author} {\bibinfo {author} {\bibfnamefont {Philipp}\ \bibnamefont
  {Strasberg}}, \bibinfo {author} {\bibfnamefont {Gernot}\ \bibnamefont
  {Schaller}}, \bibinfo {author} {\bibfnamefont {Tobias}\ \bibnamefont
  {Brandes}}, \ and\ \bibinfo {author} {\bibfnamefont {Massimiliano}\
  \bibnamefont {Esposito}},\ }\bibfield  {title} {\enquote {\bibinfo {title}
  {{Quantum and Information Thermodynamics: A Unifying Framework based on
  Repeated Interactions}},}\ }\href {\doibase 10.1103/PhysRevX.7.021003}
  {\bibfield  {journal} {\bibinfo  {journal} {Phys. Rev. X}\ }\textbf {\bibinfo
  {volume} {7}},\ \bibinfo {pages} {021003} (\bibinfo {year} {2017})},\ \Eprint
  {http://arxiv.org/abs/1610.01829} {arXiv:1610.01829} \BibitemShut {NoStop}%
\bibitem [{\citenamefont {Rodrigues}\ \emph {et~al.}(2019)\citenamefont
  {Rodrigues}, \citenamefont {{De Chiara}}, \citenamefont {Paternostro},\ and\
  \citenamefont {Landi}}]{Rodrigues2019}%
  \BibitemOpen
  \bibfield  {author} {\bibinfo {author} {\bibfnamefont {Franklin L.~S.}\
  \bibnamefont {Rodrigues}}, \bibinfo {author} {\bibfnamefont {Gabriele}\
  \bibnamefont {{De Chiara}}}, \bibinfo {author} {\bibfnamefont {Mauro}\
  \bibnamefont {Paternostro}}, \ and\ \bibinfo {author} {\bibfnamefont
  {Gabriel~T.}\ \bibnamefont {Landi}},\ }\bibfield  {title} {\enquote {\bibinfo
  {title} {{Thermodynamics of weakly coherent collisional models}},}\ }\href
  {\doibase 10.1103/PhysRevLett.123.140601} {\bibfield  {journal} {\bibinfo
  {journal} {Phys. Rev. Lett.}\ }\textbf {\bibinfo {volume} {123}},\ \bibinfo
  {pages} {140601} (\bibinfo {year} {2019})},\ \Eprint
  {http://arxiv.org/abs/1906.08203} {arXiv:1906.08203} \BibitemShut {NoStop}%
\bibitem [{\citenamefont {Barra}(2015)}]{Barra2015}%
  \BibitemOpen
  \bibfield  {author} {\bibinfo {author} {\bibfnamefont {Felipe}\ \bibnamefont
  {Barra}},\ }\bibfield  {title} {\enquote {\bibinfo {title} {{The
  thermodynamic cost of driving quantum systems by their boundaries}},}\ }\href
  {\doibase 10.1038/srep14873} {\bibfield  {journal} {\bibinfo  {journal} {Sci.
  Rep.}\ }\textbf {\bibinfo {volume} {5}},\ \bibinfo {pages} {14873} (\bibinfo
  {year} {2015})},\ \Eprint {http://arxiv.org/abs/1509.04223}
  {arXiv:1509.04223} \BibitemShut {NoStop}%
\bibitem [{\citenamefont {Pereira}(2018)}]{Pereira2018}%
  \BibitemOpen
  \bibfield  {author} {\bibinfo {author} {\bibfnamefont {Emmanuel}\
  \bibnamefont {Pereira}},\ }\bibfield  {title} {\enquote {\bibinfo {title}
  {{Heat, work, and energy currents in the boundary-driven XXZ spin chain}},}\
  }\href {\doibase 10.1103/PhysRevE.97.022115} {\bibfield  {journal} {\bibinfo
  {journal} {Phys. Rev. E}\ }\textbf {\bibinfo {volume} {97}},\ \bibinfo
  {pages} {022115} (\bibinfo {year} {2018})}\BibitemShut {NoStop}%
\bibitem [{\citenamefont {Ciccarello}(2017)}]{Ciccarello2017}%
  \BibitemOpen
  \bibfield  {author} {\bibinfo {author} {\bibfnamefont {Francesco}\
  \bibnamefont {Ciccarello}},\ }\bibfield  {title} {\enquote {\bibinfo {title}
  {{Collision models in quantum optics}},}\ }\href {\doibase
  10.1515/qmetro-2017-0007} {\bibfield  {journal} {\bibinfo  {journal} {Quantum
  Measurements and Quantum Metrology}\ }\textbf {\bibinfo {volume} {4}},\
  \bibinfo {pages} {53--63} (\bibinfo {year} {2017})},\ \Eprint
  {http://arxiv.org/abs/1712.04994} {arXiv:1712.04994} \BibitemShut {NoStop}%
\bibitem [{\citenamefont {Gross}\ \emph {et~al.}(2018)\citenamefont {Gross},
  \citenamefont {Caves}, \citenamefont {Milburn},\ and\ \citenamefont
  {Combes}}]{Gross2017a}%
  \BibitemOpen
  \bibfield  {author} {\bibinfo {author} {\bibfnamefont {Jonathan~A.}\
  \bibnamefont {Gross}}, \bibinfo {author} {\bibfnamefont {Carlton~M.}\
  \bibnamefont {Caves}}, \bibinfo {author} {\bibfnamefont {Gerard~J.}\
  \bibnamefont {Milburn}}, \ and\ \bibinfo {author} {\bibfnamefont {Joshua}\
  \bibnamefont {Combes}},\ }\bibfield  {title} {\enquote {\bibinfo {title}
  {{Qubit models of weak continuous measurements: markovian conditional and
  open-system dynamics}},}\ }\href {\doibase 10.1088/2058-9565/aaa39f}
  {\bibfield  {journal} {\bibinfo  {journal} {Quantum Sci. Technol.}\ }\textbf
  {\bibinfo {volume} {3}},\ \bibinfo {pages} {024005} (\bibinfo {year}
  {2018})},\ \Eprint {http://arxiv.org/abs/1710.09523} {arXiv:1710.09523}
  \BibitemShut {NoStop}%
\bibitem [{Note1()}]{Note1}%
  \BibitemOpen
  \bibinfo {note} {The code can be downloaded \protect \href
  {https://www.wolframcloud.com/obj/gtlandi/Published/CM2_qubit_models.nb}{here}.}\BibitemShut
  {Stop}%
\bibitem [{\citenamefont {Darwiche}(2009)}]{Darwiche2009}%
  \BibitemOpen
  \bibfield  {author} {\bibinfo {author} {\bibfnamefont {A}~\bibnamefont
  {Darwiche}},\ }\href@noop {} {\emph {\bibinfo {title} {{Modeling and
  Reasoning with Bayesian Networks}}}}\ (\bibinfo  {publisher} {Cambridge
  University Press},\ \bibinfo {address} {Cambridge},\ \bibinfo {year}
  {2009})\BibitemShut {NoStop}%
\bibitem [{\citenamefont {Neapolitan}(2003)}]{Neapolitan2003}%
  \BibitemOpen
  \bibfield  {author} {\bibinfo {author} {\bibfnamefont {R.~E.}\ \bibnamefont
  {Neapolitan}},\ }\href@noop {} {\emph {\bibinfo {title} {{Learning Bayesian
  Networks,}}}}\ (\bibinfo  {publisher} {Prentice-Hall},\ \bibinfo {address}
  {Upper Saddle River},\ \bibinfo {year} {2003})\BibitemShut {NoStop}%
\bibitem [{\citenamefont {Holevo}(1973)}]{Holevo1973}%
  \BibitemOpen
  \bibfield  {author} {\bibinfo {author} {\bibfnamefont {A.~S.}\ \bibnamefont
  {Holevo}},\ }\bibfield  {title} {\enquote {\bibinfo {title} {{Bounds for the
  Quantity of Information Transmitted by a Quantum Communication Channel}},}\
  }\href@noop {} {\bibfield  {journal} {\bibinfo  {journal} {Problems of
  Information Transmission}\ }\textbf {\bibinfo {volume} {9}},\ \bibinfo
  {pages} {177--183} (\bibinfo {year} {1973})}\BibitemShut {NoStop}%
\bibitem [{\citenamefont {Nielsen}\ and\ \citenamefont
  {Chuang}(2000)}]{Nielsen}%
  \BibitemOpen
  \bibfield  {author} {\bibinfo {author} {\bibfnamefont {M~A}\ \bibnamefont
  {Nielsen}}\ and\ \bibinfo {author} {\bibfnamefont {I~L}\ \bibnamefont
  {Chuang}},\ }\href@noop {} {\emph {\bibinfo {title} {{Quantum Computation and
  Quantum Information}}}}\ (\bibinfo  {publisher} {Cambridge University
  Press},\ \bibinfo {year} {2000})\BibitemShut {NoStop}%
\bibitem [{\citenamefont {Ficheux}\ \emph {et~al.}(2018)\citenamefont
  {Ficheux}, \citenamefont {Jezouin}, \citenamefont {Leghtas},\ and\
  \citenamefont {Huard}}]{Ficheux2018}%
  \BibitemOpen
  \bibfield  {author} {\bibinfo {author} {\bibfnamefont {Q.}~\bibnamefont
  {Ficheux}}, \bibinfo {author} {\bibfnamefont {S.}~\bibnamefont {Jezouin}},
  \bibinfo {author} {\bibfnamefont {Z.}~\bibnamefont {Leghtas}}, \ and\
  \bibinfo {author} {\bibfnamefont {B.}~\bibnamefont {Huard}},\ }\bibfield
  {title} {\enquote {\bibinfo {title} {{Dynamics of a qubit while
  simultaneously monitoring its relaxation and dephasing}},}\ }\href {\doibase
  10.1038/s41467-018-04372-9} {\bibfield  {journal} {\bibinfo  {journal}
  {Nature Communications}\ }\textbf {\bibinfo {volume} {9}},\ \bibinfo {pages}
  {1--6} (\bibinfo {year} {2018})},\ \Eprint {http://arxiv.org/abs/1711.01208}
  {arXiv:1711.01208} \BibitemShut {NoStop}%
\bibitem [{\citenamefont {Fermi}(1956)}]{Fermi1956}%
  \BibitemOpen
  \bibfield  {author} {\bibinfo {author} {\bibfnamefont {Enrico}\ \bibnamefont
  {Fermi}},\ }\href@noop {} {\emph {\bibinfo {title} {{Thermodynamics}}}}\
  (\bibinfo  {publisher} {Dover Publications Inc.},\ \bibinfo {year} {1956})\
  p.\ \bibinfo {pages} {160}\BibitemShut {NoStop}%
\bibitem [{\citenamefont {Esposito}\ \emph {et~al.}(2010)\citenamefont
  {Esposito}, \citenamefont {Lindenberg},\ and\ \citenamefont {{Van den
  Broeck}}}]{Esposito2010a}%
  \BibitemOpen
  \bibfield  {author} {\bibinfo {author} {\bibfnamefont {Massimiliano}\
  \bibnamefont {Esposito}}, \bibinfo {author} {\bibfnamefont {Katja}\
  \bibnamefont {Lindenberg}}, \ and\ \bibinfo {author} {\bibfnamefont
  {Christian}\ \bibnamefont {{Van den Broeck}}},\ }\bibfield  {title} {\enquote
  {\bibinfo {title} {{Entropy production as correlation between system and
  reservoir}},}\ }\href {\doibase 10.1088/1367-2630/12/1/013013} {\bibfield
  {journal} {\bibinfo  {journal} {New J. Phys.}\ }\textbf {\bibinfo {volume}
  {12}},\ \bibinfo {pages} {013013} (\bibinfo {year} {2010})},\ \Eprint
  {http://arxiv.org/abs/0908.1125} {arXiv:0908.1125} \BibitemShut {NoStop}%
\bibitem [{\citenamefont {Manzano}\ \emph {et~al.}(2018)\citenamefont
  {Manzano}, \citenamefont {Horowitz},\ and\ \citenamefont
  {Parrondo}}]{Manzano2017a}%
  \BibitemOpen
  \bibfield  {author} {\bibinfo {author} {\bibfnamefont {Gonzalo}\ \bibnamefont
  {Manzano}}, \bibinfo {author} {\bibfnamefont {Jordan~M.}\ \bibnamefont
  {Horowitz}}, \ and\ \bibinfo {author} {\bibfnamefont {Juan M.~R.}\
  \bibnamefont {Parrondo}},\ }\bibfield  {title} {\enquote {\bibinfo {title}
  {{Quantum fluctuation theorems for arbitrary environments: adiabatic and
  non-adiabatic entropy production}},}\ }\href
  {http://arxiv.org/abs/1710.00054} {\bibfield  {journal} {\bibinfo  {journal}
  {Phys. Rev. X}\ }\textbf {\bibinfo {volume} {8}},\ \bibinfo {pages} {031037}
  (\bibinfo {year} {2018})},\ \Eprint {http://arxiv.org/abs/1710.00054}
  {arXiv:1710.00054} \BibitemShut {NoStop}%
\bibitem [{\citenamefont {Breuer}(2003)}]{Breuer2003}%
  \BibitemOpen
  \bibfield  {author} {\bibinfo {author} {\bibfnamefont {Heinz~Peter}\
  \bibnamefont {Breuer}},\ }\bibfield  {title} {\enquote {\bibinfo {title}
  {{Quantum jumps and entropy production}},}\ }\href {\doibase
  10.1103/PhysRevA.68.032105} {\bibfield  {journal} {\bibinfo  {journal} {Phys.
  Rev. A}\ }\textbf {\bibinfo {volume} {68}},\ \bibinfo {pages} {032105}
  (\bibinfo {year} {2003})},\ \Eprint {http://arxiv.org/abs/0306047}
  {arXiv:0306047 [quant-ph]} \BibitemShut {NoStop}%
\bibitem [{\citenamefont {Landi}\ \emph {et~al.}(2021)\citenamefont {Landi},
  \citenamefont {Paternostro},\ and\ \citenamefont {Belenchia}}]{tofollow}%
  \BibitemOpen
  \bibfield  {author} {\bibinfo {author} {\bibfnamefont {G.~T.}\ \bibnamefont
  {Landi}}, \bibinfo {author} {\bibfnamefont {M.}~\bibnamefont {Paternostro}},
  \ and\ \bibinfo {author} {\bibfnamefont {A.}~\bibnamefont {Belenchia}},\
  }\href@noop {} {\enquote {\bibinfo {title} {{Informational steady-states and
  conditional entropy production in continuously monitored systems: the
  continuous-variable scenario}},}\ } (\bibinfo {year} {2021}),\ \bibinfo
  {note} {to appear}\BibitemShut {NoStop}%
\bibitem [{\citenamefont {Peebles}(1993)}]{Peebles1993}%
  \BibitemOpen
  \bibfield  {author} {\bibinfo {author} {\bibfnamefont {P.~Z.}\ \bibnamefont
  {Peebles}},\ }\href@noop {} {\emph {\bibinfo {title} {{Probability, random
  variables, and random signal principles}}}},\ \bibinfo {edition} {3rd}\ ed.\
  (\bibinfo  {publisher} {McGraw-Hill},\ \bibinfo {year} {1993})\ p.\ \bibinfo
  {pages} {448}\BibitemShut {NoStop}%
\bibitem [{\citenamefont {Chiribella}\ \emph {et~al.}(2008)\citenamefont
  {Chiribella}, \citenamefont {D'Ariano},\ and\ \citenamefont
  {Perinotti}}]{Chiribella2008a}%
  \BibitemOpen
  \bibfield  {author} {\bibinfo {author} {\bibfnamefont {G.}~\bibnamefont
  {Chiribella}}, \bibinfo {author} {\bibfnamefont {G.~M.}\ \bibnamefont
  {D'Ariano}}, \ and\ \bibinfo {author} {\bibfnamefont {P.}~\bibnamefont
  {Perinotti}},\ }\bibfield  {title} {\enquote {\bibinfo {title} {{Quantum
  Circuit Architecture}},}\ }\href {\doibase 10.1103/PhysRevLett.101.060401}
  {\bibfield  {journal} {\bibinfo  {journal} {Phys. Rev. Lett.}\ }\textbf
  {\bibinfo {volume} {101}},\ \bibinfo {pages} {060401} (\bibinfo {year}
  {2008})},\ \Eprint {http://arxiv.org/abs/0712.1325} {arXiv:0712.1325}
  \BibitemShut {NoStop}%
\bibitem [{\citenamefont {Pollock}\ \emph
  {et~al.}(2018{\natexlab{a}})\citenamefont {Pollock}, \citenamefont
  {Rodr{\'{i}}guez-Rosario}, \citenamefont {Frauenheim}, \citenamefont
  {Paternostro},\ and\ \citenamefont {Modi}}]{Pollock2018b}%
  \BibitemOpen
  \bibfield  {author} {\bibinfo {author} {\bibfnamefont {Felix~A.}\
  \bibnamefont {Pollock}}, \bibinfo {author} {\bibfnamefont {C{\'{e}}sar}\
  \bibnamefont {Rodr{\'{i}}guez-Rosario}}, \bibinfo {author} {\bibfnamefont
  {Thomas}\ \bibnamefont {Frauenheim}}, \bibinfo {author} {\bibfnamefont
  {Mauro}\ \bibnamefont {Paternostro}}, \ and\ \bibinfo {author} {\bibfnamefont
  {Kavan}\ \bibnamefont {Modi}},\ }\bibfield  {title} {\enquote {\bibinfo
  {title} {{Non-Markovian quantum processes: Complete framework and efficient
  characterization}},}\ }\href {\doibase 10.1103/PhysRevA.97.012127} {\bibfield
   {journal} {\bibinfo  {journal} {Phys. Rev. A}\ }\textbf {\bibinfo {volume}
  {97}},\ \bibinfo {pages} {012127} (\bibinfo {year} {2018}{\natexlab{a}})},\
  \Eprint {http://arxiv.org/abs/1512.00589} {arXiv:1512.00589} \BibitemShut
  {NoStop}%
\bibitem [{\citenamefont {Pollock}\ \emph
  {et~al.}(2018{\natexlab{b}})\citenamefont {Pollock}, \citenamefont
  {Rodr{\'{i}}guez-Rosario}, \citenamefont {Frauenheim}, \citenamefont
  {Paternostro},\ and\ \citenamefont {Modi}}]{Pollock2018}%
  \BibitemOpen
  \bibfield  {author} {\bibinfo {author} {\bibfnamefont {Felix~A.}\
  \bibnamefont {Pollock}}, \bibinfo {author} {\bibfnamefont {C{\'{e}}sar}\
  \bibnamefont {Rodr{\'{i}}guez-Rosario}}, \bibinfo {author} {\bibfnamefont
  {Thomas}\ \bibnamefont {Frauenheim}}, \bibinfo {author} {\bibfnamefont
  {Mauro}\ \bibnamefont {Paternostro}}, \ and\ \bibinfo {author} {\bibfnamefont
  {Kavan}\ \bibnamefont {Modi}},\ }\bibfield  {title} {\enquote {\bibinfo
  {title} {{Operational Markov Condition for Quantum Processes}},}\ }\href
  {\doibase 10.1103/PhysRevLett.120.040405} {\bibfield  {journal} {\bibinfo
  {journal} {Phys. Rev. Lett.}\ }\textbf {\bibinfo {volume} {120}},\ \bibinfo
  {pages} {040405} (\bibinfo {year} {2018}{\natexlab{b}})},\ \Eprint
  {http://arxiv.org/abs/1801.09811} {arXiv:1801.09811} \BibitemShut {NoStop}%
\bibitem [{\citenamefont {Strasberg}(2019)}]{Strasberg2019c}%
  \BibitemOpen
  \bibfield  {author} {\bibinfo {author} {\bibfnamefont {Philipp}\ \bibnamefont
  {Strasberg}},\ }\bibfield  {title} {\enquote {\bibinfo {title} {{Operational
  approach to quantum stochastic thermodynamics}},}\ }\href {\doibase
  10.1103/PhysRevE.100.022127} {\bibfield  {journal} {\bibinfo  {journal}
  {Phys. Rev. E}\ }\textbf {\bibinfo {volume} {100}},\ \bibinfo {pages}
  {022127} (\bibinfo {year} {2019})},\ \Eprint
  {http://arxiv.org/abs/1810.00698} {arXiv:1810.00698} \BibitemShut {NoStop}%
\bibitem [{\citenamefont {Alonso}\ \emph {et~al.}(2016)\citenamefont {Alonso},
  \citenamefont {Lutz},\ and\ \citenamefont {Romito}}]{Alonso2016}%
  \BibitemOpen
  \bibfield  {author} {\bibinfo {author} {\bibfnamefont {Jose~Joaquin}\
  \bibnamefont {Alonso}}, \bibinfo {author} {\bibfnamefont {Eric}\ \bibnamefont
  {Lutz}}, \ and\ \bibinfo {author} {\bibfnamefont {Alessandro}\ \bibnamefont
  {Romito}},\ }\bibfield  {title} {\enquote {\bibinfo {title} {{Thermodynamics
  of Weakly Measured Quantum Systems}},}\ }\href {\doibase
  10.1103/PhysRevLett.116.080403} {\bibfield  {journal} {\bibinfo  {journal}
  {Phys. Rev. Lett.}\ }\textbf {\bibinfo {volume} {116}},\ \bibinfo {pages}
  {080403} (\bibinfo {year} {2016})},\ \Eprint
  {http://arxiv.org/abs/1508.00438} {arXiv:1508.00438} \BibitemShut {NoStop}%
\bibitem [{\citenamefont {Naghiloo}\ \emph {et~al.}(2020)\citenamefont
  {Naghiloo}, \citenamefont {Tan}, \citenamefont {Harrington}, \citenamefont
  {Alonso}, \citenamefont {Lutz}, \citenamefont {Romito},\ and\ \citenamefont
  {Murch}}]{Naghiloo2020}%
  \BibitemOpen
  \bibfield  {author} {\bibinfo {author} {\bibfnamefont {M.}~\bibnamefont
  {Naghiloo}}, \bibinfo {author} {\bibfnamefont {D.}~\bibnamefont {Tan}},
  \bibinfo {author} {\bibfnamefont {P.~M.}\ \bibnamefont {Harrington}},
  \bibinfo {author} {\bibfnamefont {J.~J.}\ \bibnamefont {Alonso}}, \bibinfo
  {author} {\bibfnamefont {E.}~\bibnamefont {Lutz}}, \bibinfo {author}
  {\bibfnamefont {A.}~\bibnamefont {Romito}}, \ and\ \bibinfo {author}
  {\bibfnamefont {K.~W.}\ \bibnamefont {Murch}},\ }\bibfield  {title} {\enquote
  {\bibinfo {title} {{Heat and Work Along Individual Trajectories of a Quantum
  Bit}},}\ }\href {\doibase 10.1103/PhysRevLett.124.110604} {\bibfield
  {journal} {\bibinfo  {journal} {Phys. Rev. Lett.}\ }\textbf {\bibinfo
  {volume} {124}},\ \bibinfo {pages} {110604} (\bibinfo {year} {2020})},\
  \Eprint {http://arxiv.org/abs/1703.05885} {arXiv:1703.05885} \BibitemShut
  {NoStop}%
\bibitem [{\citenamefont {Groenewold}(1971)}]{Groenewold1971}%
  \BibitemOpen
  \bibfield  {author} {\bibinfo {author} {\bibfnamefont {H.~J.}\ \bibnamefont
  {Groenewold}},\ }\bibfield  {title} {\enquote {\bibinfo {title} {{A problem
  of information gain by quantal measurements}},}\ }\href {\doibase
  10.1007/BF00815357} {\bibfield  {journal} {\bibinfo  {journal} {Int. J.
  Theor. Phys.}\ }\textbf {\bibinfo {volume} {4}},\ \bibinfo {pages} {327--338}
  (\bibinfo {year} {1971})}\BibitemShut {NoStop}%
\bibitem [{\citenamefont {Ozawa}(1986)}]{Ozawa1986}%
  \BibitemOpen
  \bibfield  {author} {\bibinfo {author} {\bibfnamefont {Masanao}\ \bibnamefont
  {Ozawa}},\ }\bibfield  {title} {\enquote {\bibinfo {title} {{On Information
  gain by quantum measurements of continuous observables}},}\ }\href {\doibase
  10.1063/1.527179} {\bibfield  {journal} {\bibinfo  {journal} {J. Math.
  Phys.}\ }\textbf {\bibinfo {volume} {27}},\ \bibinfo {pages} {759--763}
  (\bibinfo {year} {1986})}\BibitemShut {NoStop}%
\bibitem [{\citenamefont {Gardas}\ and\ \citenamefont
  {Deffner}(2018)}]{gardas2018quantum}%
  \BibitemOpen
  \bibfield  {author} {\bibinfo {author} {\bibfnamefont {Bart{\l}omiej}\
  \bibnamefont {Gardas}}\ and\ \bibinfo {author} {\bibfnamefont {Sebastian}\
  \bibnamefont {Deffner}},\ }\bibfield  {title} {\enquote {\bibinfo {title}
  {Quantum fluctuation theorem for error diagnostics in quantum annealers},}\
  }\href@noop {} {\bibfield  {journal} {\bibinfo  {journal} {Scientific
  reports}\ }\textbf {\bibinfo {volume} {8}},\ \bibinfo {pages} {1--8}
  (\bibinfo {year} {2018})}\BibitemShut {NoStop}%
\bibitem [{\citenamefont {Buffoni}\ and\ \citenamefont
  {Campisi}(2020)}]{buffoni2020thermodynamics}%
  \BibitemOpen
  \bibfield  {author} {\bibinfo {author} {\bibfnamefont {Lorenzo}\ \bibnamefont
  {Buffoni}}\ and\ \bibinfo {author} {\bibfnamefont {Michele}\ \bibnamefont
  {Campisi}},\ }\bibfield  {title} {\enquote {\bibinfo {title} {Thermodynamics
  of a quantum annealer},}\ }\href@noop {} {\bibfield  {journal} {\bibinfo
  {journal} {Quantum Science and Technology}\ }\textbf {\bibinfo {volume}
  {5}},\ \bibinfo {pages} {035013} (\bibinfo {year} {2020})}\BibitemShut
  {NoStop}%
\bibitem [{\citenamefont {Cimini}\ \emph {et~al.}(2020)\citenamefont {Cimini},
  \citenamefont {Gherardini}, \citenamefont {Barbieri}, \citenamefont
  {Gianani}, \citenamefont {Sbroscia}, \citenamefont {Buffoni}, \citenamefont
  {Paternostro},\ and\ \citenamefont {Caruso}}]{cimini2020experimental}%
  \BibitemOpen
  \bibfield  {author} {\bibinfo {author} {\bibfnamefont {Valeria}\ \bibnamefont
  {Cimini}}, \bibinfo {author} {\bibfnamefont {Stefano}\ \bibnamefont
  {Gherardini}}, \bibinfo {author} {\bibfnamefont {Marco}\ \bibnamefont
  {Barbieri}}, \bibinfo {author} {\bibfnamefont {Ilaria}\ \bibnamefont
  {Gianani}}, \bibinfo {author} {\bibfnamefont {Marco}\ \bibnamefont
  {Sbroscia}}, \bibinfo {author} {\bibfnamefont {Lorenzo}\ \bibnamefont
  {Buffoni}}, \bibinfo {author} {\bibfnamefont {Mauro}\ \bibnamefont
  {Paternostro}}, \ and\ \bibinfo {author} {\bibfnamefont {Filippo}\
  \bibnamefont {Caruso}},\ }\bibfield  {title} {\enquote {\bibinfo {title}
  {Experimental characterization of the energetics of quantum logic gates},}\
  }\href@noop {} {\bibfield  {journal} {\bibinfo  {journal} {npj Quantum
  Information}\ }\textbf {\bibinfo {volume} {6}},\ \bibinfo {pages} {1--8}
  (\bibinfo {year} {2020})}\BibitemShut {NoStop}%
\end{thebibliography}%
%\bibliography{/Users/gtlandi/Documents/library}
\end{document}